\documentclass[12pt]{iopart}
\usepackage{mathrsfs}
\usepackage{graphicx,color}

\usepackage{iopams}
\def\ra{\rangle}
\def\la{\langle}

\def\dag{^\dagger}

\def\Oh{{\hat \Omega}}

\def\vs{{\vec s}}

\def\be{\beta}
\def\al{\alpha}

\def\del{\delta}

\def\bege{\begin{equation}}
\def\ende{\end{equation}}
\def\begen{\begin{eqnarray}}
\def\enden{\end{eqnarray}}

\newcommand{\vectorize}[2]{\begin{array}{c} #1 \\ #2 \end{array}}
\newcommand{\shelf}[3]{\begin{array}{c|c} #1 & \begin{array}{c} #2 \\ #3 \end{array} \end{array}}
\newcommand{\ssf}[1]{\sf #1}

\begin{document}

\title[]{Entanglement in valence-bond-solid states on symmetric graphs}

\author{Hosho Katsura${}^{1}$, Naoki Kawashima${}^2$, Anatol N. Kirillov${}^3$, Vladimir E. Korepin${}^4$, Shu Tanaka${}^{2}$}

\address{${}^1$ Kavli Institute for Theoretical Physics, University of California,
Santa Barbara, CA 93106, USA}
\address{${}^2$ Institute for Solid State Physics, The University of Tokyo, 5-1-5 Kashiwanoha, Kashiwa, Chiba, 
277-8581, Japan}
\address{${}^3$ Research Institute for Mathematical Sciences,
Kyoto University, Sakyo-ku,
Kyoto, 606-8502, Japan}
\address{${}^4$ C.N. Yang Institute for Theoretical Physics
State University of New York at Stony Brook
Stony Brook, NY 11794-3840}
\ead{
\mailto{katsura@kitp.ucsb.edu},
\mailto{kawashima@issp.u-tokyo.ac.jp}
\mailto{kirillov@kurims.kyoto-u.ac.jp}, 
\mailto{korepin@max2.physics.sunysb.edu} 
\mailto{shu-t@issp.u-tokyo.ac.jp}
}
\begin{abstract}
We study quantum entanglement in the ground state of the Affleck-Kennedy-Lieb-Tasaki (AKLT) model defined on two-dimensional graphs with reflection and/or inversion symmetry. The ground state of this spin model is known as the valence-bond-solid state. 
We investigate the properties of reduced density matrix of a subsystem which is a mirror image of the other one. 
Thanks to the reflection symmetry, the eigenvalues of the reduced density matrix can be obtained by numerically 
diagonalizing a real symmetric matrix whose elements are calculated by Monte Carlo integration. 
We calculate the von Neumann entropy of the reduced density matrix. 
The obtained results indicate that there is some deviation from the naive expectation that the von Neumann entropy per 
valence bond on the boundary between the subsystems is $\ln 2$. 
This deviation is interpreted in terms of the hidden spin chain along the boundary between the subsystems. 
In some cases where graphs are on ladders, the numerical results are analytically or algebraically confirmed. 
\end{abstract}

\submitto{\JPA}
\maketitle

\section{Introduction}
In recent years, the study of entanglement in quantum many-body systems has become a common issue in condensed matter physics, statistical mechanics, and quantum information theory. 
Entanglement is a purely quantum mechanical phenomenon in which quantum systems are linked together even when they are spatially separated so that one system cannot be correctly described without full mention of its counterpart. 
This concept was first introduced by Schr\"odinger in~\cite{Schroedinger}. 
From the viewpoint of quantum information science, entanglement is a fundamental measure of how much quantum effects we can observe and use to control one quantum system by another, and it is the primary resource for quantum information processing and communication~\cite{Nielsen_Chuang, Llyoid, Bennet_DiVincenzo}. 
On the other hand, in condensed matter physics and statistical mechanics, the concept of quantum entanglement has recently been used to investigate quantum phase transitions~\cite{Vidal} (see~\cite{Amico_review} for a review), topological order~\cite{Kitaev_Preskill, Levine_Wen, Ryu_Hatsugai}, and macroscopic properties of solids~\cite{Ghosh}. 
A fundamental and practical problem common in all the areas is how to detect entanglement~\cite{Toth} and quantify the degree of it. 
A general approach is to study the reduced density matrix of a certain subsystem in an entangled state. 
The concept of the reduced density matrix was first introduced by Dirac in~\cite{Dirac}. 
There are many kinds of measures of entanglement related to the reduced density matrix (see~\cite{Horodecki4} and references therein). 
Among them, one of the most popular characteristic functions is the von Neumann entropy (entanglement entropy) of a reduced density matrix~\cite{Bennet, Vedral}. It serves as the measure of entanglement for a pure state. 

Generally speaking, to elucidate quantum entanglement in many-body systems is a formidable task since we need a precise description of  many-body quantum states. 
So far, much insight has been obtained by studying exactly solvable models in which there is a possibility to obtain the reduced density matrix and calculate the von Neumann entropy exactly or approximately. 
Representative models are the harmonic models ~\cite{Audenaert_Eisert, Cramer_Eisert}, 
the XY spin chain~\cite{Jin_Korepin_04, Peschel, Its_Jin_Korepin_05, Franchini_Its_Jin_Korepin_07, Franchini_Its_Korepin_08, Korepin_Its_09}, 
the XXZ spin chain~\cite{Gu_Lin_Li, Sato_Shiroishi, Nienhuis_Calabrese}, 
the Calogero-Sutherland model~\cite{Katsura_Hatsuda}, 
one-dimensional critical models~\cite{Korepin_PRL},  
conformal field theories~\cite{Calabrese_Cardy1, Calabrese_Cardy2}, 
the Affleck-Kennedy-Lieb-Tasaki (AKLT) model~\cite{Fan_Korepin1, Hirano_Hatsugai, Katsura_Hirano_Hatsugai}, 
the Kitaev model~\cite{Zanardi_Kitaev}, and 
the quantum dimer (Rokhsar-Kivelson) model~\cite{Furukawa_Misguich}. 
Among them, the AKLT model~\cite{AKLT_PRL, AKLT_CMP} has some important meanings since this model is directly related to one of the schemes of quantum computation, measurement based quantum computation~\cite{Verstraete_Cirac, Brennen_Miyake}. 
Historically, the AKLT model and its exact ground state known as the valence-bond-solid (VBS) state were proposed to understand ground-state properties of the Haldane gap systems~\cite{Haldane_PLA, Haldane_PRL}. 
The authors (AKLT) rigorously proved that the VBS state is a unique ground state and there is an energy gap immediately above the ground state. Exponential decay of correlation functions in the VBS ground state has also been shown. There is an intimate relation between the VBS states and fractional quantum Hall states which has been revealed by using a spin coherent state representation~\cite{Arovas_Auerbach_Haldane}. A hidden topological order in the VBS state is characterized by string order parameters~\cite{Nijs_Rommelse} and existence of edge states which are degenerate ground states in a chain with open boundary conditions~\cite{Kennedy_Tasaki_CMP, Oshikawa_JPC}. The VBS picture and the edge states were also experimentally supported by electron spin resonance (ESR) in the $S=1$ Haldane chain Ni(C$_2$H$_8$N$_2$)$_2$NO$_2$(ClO$_4$) (NENP)~\cite{Hagiwara_Katsumata, Glarum}. 

Although most of the exactly solvable models are limited to one dimension, the AKLT model and VBS ground state can be generalized to higher dimensions and/or inhomogeneous (non-translational invariant) lattices~\cite{AKLT_CMP, Kennedy_Lieb_Tasaki, Kirillov_Korepin1, Kirillov_Korepin2, Freitag2}. 
This feature is especially important in connection with the measurement based quantum computation as we discussed later.
For studying nature of the VBS states, the non-trivial observation has been done by Kl\"umper {\it et al}., through the $q$-deformation of the AKLT model~\cite{Kluemper0, Kluemper1, Kluemper2}. 
They found that the one-dimensional (1d) spin-1 VBS state can be expressed in a form of the matrix product state (MPS). This representation gives an efficient method to calculate correlation functions. 
Mathematical theory and generalization of the VBS states, i.e.,  finitely correlated states (FCS), were essentially developed by Fannes {\it et al}., in~\cite{Fannes0, Fannes1,Fannes2, Fannes3}. 
Another interesting generalization of the VBS state is to replace SU(2) spins with other Lie algebra generators. 
In this direction, there have been constructed many versions of the AKLT model such as 
SU(3)~\cite{Greiter_Rachel_Schuricht} and SU($n$) modes~\cite{Greiter_Rachel, Rachel_Thomale, Katsura_Hirano_Korepin}, 
SU($2n$)-extended model~\cite{XVBS}, symplectic model~\cite{Schuricht_Rachel}, 
SO(5) model~\cite{Scalapino_ladder}, SO($2S+1$) model~\cite{Zhang_Xiang, Tu_Zhang_Xiang}, 
supersymmetric model~\cite{Hasebe}, and  
several $q$-deformations~\cite{Batchelor0, Batchelor1, Koo_Saleur, Fannes_q, Gils, Motegi}. 

The VBS state has recently been attracting renewed interest from the viewpoint of quantum information theory. 
In particular, it has been proposed that the VBS state can be used as a resource state in measurement-based quantum computation (MQC). 
MQC is one of the schemes of quantum computation in which we can perform universal quantum computation using only measurements as computational steps (see~\cite{Jozsa} for an introduction). 
There are two kinds of measurement based models. One is the teleportation quantum computation (TQC)~\cite{Gottesman_MQC, Nielsen_MQC, Leung_MQC}. The other is the so-called one-way quantum computation (1WQC) or cluster state computation~\cite{Briegel1, Briegel2, Briegel3, Nielsen_1WQC}. 
In this model, any quantum gate array can be implemented as a pattern of single-qubit measurement on a highly entangled cluster state. 
Note that 1WQC with one-dimensional cluster states is insufficient for universal quantum computation and thus two-dimensional cluster states are inevitably needed. It has then been proved in~\cite{Verstraete_Cirac} by Verstraete and Cirac that TQC and 1WQC are equivalent using the VBS picture. 
Gross, Eisert and collaborators have introduced novel schemes for MQC based on finitely correlated or tensor network state beyond the cluster state~\cite{Gross_Eisert1, Gross_Eisert2}. 
Another recent important development by Brennen and Miyake is that the VBS state (more precisely dynamically coupled AKLT chains) can be used as a resource state in MQC instead of the cluster state~\cite{Brennen_Miyake}. 
The VBS state is more advantageous to 1WQC than the cluster state since it is robust against noise due to the energy gap. 
Implementations of the AKLT Hamiltonian using bosonic atoms~\cite{Yip, Garcia_Ripoll} or spin-1 polar molecules~\cite{Brennen_Zoller} in optical lattices have also been proposed. 

In this paper, motivated by considerable current interest from several viewpoints, we study entanglement properties of VBS states on two-dimensional (2d) graphs which is relevant to MQC. 
Before discussing the 2d VBS states, let us briefly summarize the obtained results in 1d VBS states. 
The reduced density matrix and entanglement properties in the 1d spin-1 VBS state was first studied in~\cite{Fan_Korepin1}. 
The authors found that the von Neumann entropy of the block with the rest of the chain approaches a constant value exponentially fast, as the size of the block increases. The constant value is given by $2 \ln 2$. Then, the 1d integer-$S$ VBS states were studied in~\cite{Katsura_Hirano_Hatsugai}. 
In this case, the von Neumann entropy of the large block with the rest is given by $2\ln (S+1)$ and this value is interpreted in terms of the edge states, the degenerate ground states of the open AKLT chain. This picture leads to an idea that the reduced density matrix of the block in the AKLT chain is {\it exactly} spanned by the degenerate ground states of the block Hamiltonian (see Eq. (\ref{block_Ham}) for the definition). This idea has been confirmed in the series of papers~\cite{Katsura_Hirano_Korepin, XKHK1, XKHK2} and was summarized in the review~\cite{Korepin_Xu_review}. 
Another interesting quantities such as the entanglement length~\cite{Verstraete_Delgado_Cirac}, the geometric entanglement~\cite{Orus}, the single-copy entanglement~\cite{Hadley}, and the boundary effect~\cite{Fan_Korepin3} in the 1d spin-1 VBS state have also been studied. More generally, the entanglement of formation in the 1d FCS was estimated in~\cite{Michalakis} and an area law in 1d gapped quantum system was proven in~\cite{Hastings}. 
Compared to the 1d VBS states, few results have been obtained in 2d VBS states so far. 
Entanglement properties of the VBS state on the Cayley tree~\cite{AKLT_CMP, Niggemann_Zittarz} was studied in~\cite{Fan_Korepin2}. The authors showed that the von Neumann entropy of the block does not depend on the whole size of the system and its asymptotic value is linearly proportional to the number of valence bonds crossing the boundary. 
This gives an explicit proof of the area law in this system (see~\cite{Eisert_area_law_review} for a review of area laws and the entanglement entropy). 
General entanglement properties of the VBS state on an arbitrary graph~\cite{AKLT_CMP, Kirillov_Korepin1, Kirillov_Korepin2} have been studied in~\cite{Xu_Korepin}. 
The authors showed that the eigenspace of the reduced density matrix of the block is spanned by the degenerate ground states of the block Hamiltonian. 
The authors have shown that the von Neumann entropy of a large block with the rest approaches a value less than $\ln (deg.)$, where $(deg.)$ is the ground-state degeneracy of the block Hamiltonian. 
Although some general results for 2d VBS states have been obtained so far, a quantitative analysis of entanglement was missing since a powerful analytical method such as a transfer matrix technique does not work sufficiently in the analysis of 2d systems. 
Therefore, entanglement properties of the 2d VBS states must be analyzed with the aid of numerics.  

In this paper, we study entanglement properties of the 2d VBS states on symmetric graphs. 
Symmetries such as reflection and inversion enable us to develop an efficient method to study the reduced density matrix. 
Due to the symmetry, one can find the relation between the eigenvalues of the reduced density matrix and those of the {\it overlap} matrix (see Eqs. (\ref{overlap1}-\ref{overlap2}) for the definition). The overlap matrix is a real symmetric and its elements are given by the inner product of the degenerate ground states of the block Hamiltonian which are linearly independent but may not be orthonormal. 
The inner product can be, for example, calculated by Monte Carlo integration using the Schwinger boson representation of the VBS state. 
Even without numerical integration, one can formally write the overlap matrix as a matrix-valued correlation function in the VBS state restricted to one of the subsystems. From this, we conjecture that the von Neumann entropy per valence bond (boundary site) is strictly less than $\ln 2$ even in the 2d infinite system. A {\it holographic} interpretation of the deviation from $\ln 2$ is given in terms of the hidden spin chain along the boundary between two subsystems (reflection axis). 
To confirm this conjecture, we numerically study the eigenvalues of the overlap matrix by exact diagonalization. Since the dimension of the overlap matrix does not depend on the size of the whole system but is given by $(deg.)$, we can study relatively large systems. 
From the eigenvalues of the reduced density matrix, we calculate the von Neumann entropy. We find that the von Neumann entropy per valence bond on the boundary is strictly less than $\ln 2$ for finite size systems. We also find a function which fits the obtained numerical results well. The long-distance behavior of this function strongly supports our conjecture. 
These numerical results are analytically or algebraically confirmed for the graphs with a ladder geometry, where transfer matrix technique can be applied. 

The organization of this work is as follows. In Section 2, we review the construction of the AKLT model on an arbitrary graph and introduce the Schwinger boson representation of the VBS state.  
In section 3, we provide a general argument on the Schmidt decomposition and discuss a relation between the reduced density matrix and the overlap matrix. We show how to apply this method to study the von Neumann entropy of the VBS state. Then we make a conjecture on the von Neumann entropy in the 2d infinite system.  
In Section 4, numerical results of the VBS states on the square and hexagonal lattices are shown. 
In Section 5, the obtained numerical results are partly confirmed analytically for graphs in a ladder geometry. 
Conclusions are given in the last section. 
In Appendix, we provide a graphical interpretation of the overlap matrix in terms of closed loops and open strands. 

\section{The basic AKLT model}
\begin{figure}[tb]
\begin{center}
\vspace{.5cm}
\hspace{-.0cm}\includegraphics[width=0.5\columnwidth]{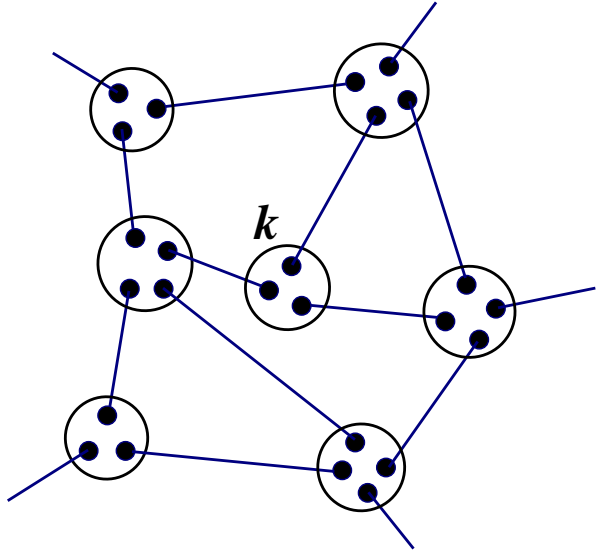}
\vspace{.0cm}
\caption{VBS state in the basic AKLT model on a graph. Circles represent vertices while solid lines represent edges. The vertex $k$ has the spin $S_k=3/2$ in this graph.}
\label{fig: VBS_basic_model}
\end{center}
\end{figure}

Let us first define the {\it basic} AKLT model on a connected graph. 
A graph consists of two types of elements, namely vertices (sites) and edges (bonds). 
As shown in Fig. \ref{fig: VBS_basic_model}, a vertex is drawn as a (large) circle and an edge is as a solid line. 
A graph is called {\it connected} if every pair of distinct vertices in the graph can be connected through some path.  Let $G=(V, E)$ be a connected graph, where $V$ and $E$ are the sets of vertices and edges, respectively. Henceforth, we assume $|V|>1$, otherwise there is no interaction between spins. 
The spin operator ${\vec S}_k$ is located at vertex $k$ and its spin value is denoted by $S_k$. 
In the basic model, we require that $S_k=\frac{1}{2}z_k$, where $z_k$ is the coordination number, i.e., the number of incident edges connected to $k$. 
To guarantee uniqueness of the ground state, this relation must be true for any vertex $k$, including boundaries.  Let us now define the spin Hamiltonian for the basic AKLT model. 
The Hamiltonian is written as a sum of interactions on all the edges: 
\bege
H= \sum_{\la k, l \ra \in E} A(k,l) \pi_{S_k+S_l} (k,l). 
\label{basic_AKLT_ham}
\ende
Here, $A(k,l)$ is an arbitrary positive real coefficient (it may depend on the edge $\la k,l \ra$) and the operator $\pi_{S_k+S_l}(k, l)$ projects the total spin on the edge $\la k,l \ra$, $\vec J_{k,l}=\vec S_k + \vec S_l$, on the subspace with the highest possible spin value $J_{k,l}=S_k+S_l$. 
Since the Hamiltonian in Eq. (\ref{basic_AKLT_ham}) is a linear combination of projection operators with positive coefficients, $H$ is positive semi-definite. 
The explicit form of the projector $\pi_{S_k+S_l}(k,l)$ in terms of $\vec S_k$ and $\vec S_l$ is given by
\bege
\pi_{S_k+S_l}(k,l)=\prod_{|S_k-S_l| \le j \le S_k+S_l-1} \frac{(\vec S_k+\vec S_l)^2-j(j+1)}{(S_k+S_l)(S_k+S_l+1)-j(j+1)}.
\label{projector_pi}
\ende
Note that one can expand $(\vec S_k+\vec S_l)^2=2 \vec S_k \cdot \vec S_l+S_k(S_k+1)+S_l(S_l+1)$ and hence the projector defined in Eq. (\ref{projector_pi}) is a polynomial in the scalar product $(\vec S_k \cdot \vec S_l)$ of the degree $2 S_{\rm min}$, where $S_{\rm min}=S_k$ is the minimum of  $S_k$ and $S_l$. As an example, consider a case where $S_k=S_l=1$, which corresponds to the $S=1$ homogeneous AKLT chain. The projector $\pi_{2}(k,l)$ is written as
\bege
\pi_2(k,l)=\frac{1}{6} (\vec S_k \cdot \vec S_l)^2 +\frac{1}{2} (\vec S_k \cdot \vec S_l)+\frac{1}{3}. 
\ende
The projectors $\pi_{3}(k,l)$ and $\pi_{4}(k,l)$ are related to the Hamiltonians on the hexagonal and the square lattices, respectively: 
\begin{eqnarray}
\pi_{3}(k,l)=\frac{1}{90}(\vec S_k \cdot \vec S_l)^3 + \frac{29}{360}(\vec S_k \cdot \vec S_l)^2 + \frac{27}{160}(\vec S_k \cdot \vec S_l) + \frac{11}{128}, \\
\pi_{4}(k,l)=\frac{1}{2520}(\vec S_k \cdot \vec S_l)^4 + \frac{1}{180}(\vec S_k \cdot \vec S_l)^3 + \frac{1}{40}(\vec S_k \cdot \vec S_l)^2 + \frac{1}{28}(\vec S_k \cdot \vec S_l).
\end{eqnarray}
In general, explicit expressions in terms of $(\vec S_k \cdot \vec S_l)$ become more complicated.

Let us now discuss the ground state of $H$ in Eq. (\ref{basic_AKLT_ham}) with condition 
\bege
S_k=\frac{1}{2}z_k. 
\ende
The ground state is unique and known as the valence-bond-solid (VBS) state. 
Although there are several possible representations for the VBS state (see~\cite{Korepin_Xu_review, Xu_Korepin}), the most convenient one for our purposes is the Schwinger boson representation~\cite{Arovas_Auerbach_Haldane, Kirillov_Korepin1, Kirillov_Korepin2}. 
In this representation, we introduce a pair of bosonic creation and annihilation operators at each vertex to realize SU(2) Lie algebra. 
We define a pair of bosonic operators $a_k$ and $b_k$ for each vertex $k$ with the canonical commutation relations:
\bege
[a_k, a^\dagger_l]=[b_k, b^\dagger_l] = \delta_{kl}
\ende
with all other commutators vanishing:
\bege
[a_k, a_l]=[b_k, b_l]=[a_k, b_l]=[a_k, b^\dagger_l]=0,~~~~~{}^\forall (k,l).
\ende
Spin operators are represented by the Schwinger bosons as
\bege
S^+_k=a^\dagger_k b_k,~~~S^-_k=b^\dagger_k a_k, ~~~ S^z_k =\frac{1}{2} (a^\dagger_k a_k-b^\dagger_k b_k). 
\ende
To reproduce the dimension of the spin-$S_k$ Hilbert space at each vertex $k$, the following constraint on the total boson occupation number is required:
\bege
a^\dagger_k a_k + b^\dagger_k b_k =2 S_k.
\ende
With this constraint in mind, the VBS ground state is written as
\bege
|{\rm VBS} \ra =\prod_{\la k,l\ra \in E} (a^\dagger_k b^\dagger_l-b^\dagger_k a^\dagger_l) |{\rm vac}\ra,
\label{vbs_gs}
\ende
where the product runs over all edges and the vacuum $|{\rm vac}\ra$ is defined as the direct product of vacuums of each vertex, i.e., 
\bege
|{\rm vac}\ra = \bigotimes_k |{\rm vac}^{[k]} \ra.
\ende
Note that $|{\rm vac}^{[k]}\ra$ is destroyed by any annihilation operators $a_k$ or $b_k$. 
Using the Schwinger boson representation, one can generalize the VBS state which is the ground state of the {\it generalized} AKLT model. 
By associating a positive integer $M_{k,l}$ to each edge $\la k,l\ra$ of $G$, the generalized VBS state is written as
\bege
|{\rm VBS}_{\rm Gen} \ra =\prod_{\la k,l\ra \in E} (a^\dagger_k b^\dagger_l-b^\dagger_k a^\dagger_l)^{M_{k,l}} |{\rm vac}\ra.
\ende
For the construction of the generalized AKLT Hamiltonian and the condition of the uniqueness of the ground state, please see original references~\cite{Kirillov_Korepin1, Kirillov_Korepin2, Korepin_Xu_review, Xu_Korepin}.
In this paper, we shall focus on the basic model, i.e., $M_{k,l}=1$ for any $\la k,l \ra$. 

Let us now prove that the VBS state in Eq. (\ref{vbs_gs}) is the zero-energy groud state of the basic AKLT Hamiltonian $H$. 
To prove this, we have only to show for any vertex $k$ and edge $\la k,l \ra$ (i) the total power of $a^\dagger_k$ and $b^\dagger_k$ is $2S_k$ so that the spin value at the vertex $k$ $S_k$, and (ii) the maximum value of the total spin of the edge $\la k,l \ra$ is $S_k+S_l-1$. 
(i) can be shown by expanding r.h.s. of Eq. (\ref{vbs_gs}) and finding that the total power of $a^\dagger_k$ and $b^\dagger_k$ is $z_k=2S_k$. 
Let us prove (ii). By counting the power of $a^\dagger$'s minus $b^\dagger$'s on the bond $\la k,l \ra$, we find that the maximal eigenvalue of $S^z_k + S^z_l$ is $S_k + S_l-1$, i.e., $J^z_{k,l} \le S_k+S_l-1$. Since the state $|{\rm VBS} \ra$ is invariant under global rotations, the maximum value of the total spin of the edge $\la k,l \ra$ is $S_k+S_l-1$. 
Therefore, from (i) and (ii), the VBS state $|{\rm VBS}\ra$ in Eq. (\ref{vbs_gs}) is a zero-energy ground state of the basic Hamiltonian $H$. It was shown in \cite{Kennedy_Lieb_Tasaki,Kirillov_Korepin1} that this ground state is unique. 

\section{Schmidt decomposition and VBS states on symmetric graphs}
\label{sec: schmidt}
In this section, we apply the Schmidt decomposition method to the VBS state on a reflection symmetric graph.
We will see that an upper bound on the von Neumann entropy can be easily obtained by this method. 
We shall start with the Schmidt decomposition for a general state. 
We follow the approach by Shi {\it et al.} in~\cite{Shi_Vidal}. 
Suppose that the state $|\Psi\ra$ of a total system is written as
\begin{equation}
|\Psi\ra=\sum_{\al} |\phi^{[A]}_\al \ra \otimes |\phi^{[B]}_\al\ra,
\label{pre_Schmidt}
\end{equation}
where $A$ and $B$ denote subsystems $A$ and $B$, respectively. We call the above decomposition {\it pre-Schmidt decomposition}. Note that the sets of states $\{|\phi^{[A]}_\al \ra \}$ and $\{|\phi^{[B]}_\al\ra \}$ are linearly independent but may not be orthonormal. In this sense, the above expression is not of the form of the Schmidt decomposition.
To Schmidt decompose the above state, we now define {\it overlap} matrices $M^{[A]}$ and $M^{[B]}$ as
\begin{eqnarray}
(M^{[A]})_{\al \be} & \equiv & \la \phi^{[A]}_\al|\phi^{[A]}_\be \ra, \label{overlap1} \\
(M^{[B]})_{\al \be} & \equiv & \la \phi^{[B]}_\al|\phi^{[B]}_\be \ra, \label{overlap2}
\end{eqnarray} 
respectively. Their spectral decompositions are given by 
\begin{eqnarray}
M^{[A]} &=& {\tilde X} D^{[A]} {\tilde X}\dag, ~~~~~
({\tilde X}\dag {\tilde X}={\tilde X}{\tilde X}\dag=1), \\
M^{[B]} &=& {\tilde Y} D^{[B]} {\tilde Y}\dag, ~~~~~
({\tilde Y}\dag {\tilde Y}={\tilde Y}{\tilde Y}\dag=1),
\end{eqnarray}
where $D^{[A]}$ and $D^{[B]}$ are diagonal matrices, i.e., $(D^{[a]})_{\tau \tau'}=\del_{\tau \tau'} d^{[a]}_\tau$ ($a=A$ or $B$).
Using ${\tilde X}$ and ${\tilde Y}$, one can obtain the orthonormal sets in $A$ and $B$ as
\begin{equation}
|e_\tau\ra = \frac{1}{\sqrt{d^{[A]}_\tau}}\sum_\al ({\tilde X}\dag)_{\tau \al}|\phi^{[A]}_\al \ra,~~~~~
|f_\eta\ra = \frac{1}{\sqrt{d^{[B]}_\eta}}\sum_\al ({\tilde Y}\dag)_{\eta \al}|\phi^{[B]}_\al \ra.
\end{equation}
Then, Eq. (\ref{pre_Schmidt}) can be written as
\begin{equation}
|\Psi\ra = \sum_{\tau \eta}\sqrt{d^{[A]}_\tau d^{[B]}_\eta} ({\tilde X}^{\rm T}{\tilde Y})_{\tau \eta} |e_\tau\ra \otimes |f_\eta\ra,
\label{pre_Schmidt2}
\end{equation}
where $^{\rm T}$ denotes matrix transpose. So far, we have considered a general case. Let us now focus on the special situation where $M^{[A]}$ and $M^{[B]}$ are the same, i.e., $M^{[A]}=M^{[B]}=M$. Furthermore, if $M$ is a real symmetric matrix, it can be diagonalized by some orthogonal matrix $O$ as $M=O D O^{\rm T}$. In such a case, Eq. (\ref{pre_Schmidt2}) becomes a simpler form:
\begin{equation}
|\Psi\ra = \sum_\tau d_\tau |e_\tau\ra \otimes |f_\tau\ra.
\end{equation}
Note that $d_\tau = d^{[A]}_\tau=d^{[B]}_\tau$. One may think that the above situation is not generic. However, as we will see later, it is enough for our purpose to restrict ourselves to such a special case.  
Let us now consider the density matrix for the total system consisting of $A$ and $B$. It is defined by
$\rho=|\Psi\ra \la \Psi|/\la \Psi|\Psi\ra$. Then the reduced density matrix for $A$ is given by $\rho_A=\tr_B \rho$. In our case, it is given by
\begin{eqnarray}
\rho_A &=& \sum_\eta \la f_\eta| \rho |f_\eta\ra \nonumber \\
&=& \sum_\eta \sum_{\tau_1} \sum_{\tau_2}d_{\tau_1}d_{\tau_2} \la f_\eta |
(|e_{\tau_1} \ra \otimes |f_{\tau_1}\ra \la e_{\tau_2}|\otimes \la f_{\tau_2}|)|f_\eta\ra/ \la \Psi|\Psi\ra \nonumber \\
&=& \frac{\sum_\tau d^2_\tau |e_\tau \ra \la e_\tau|}{\sum_\tau d^2_\tau}.
\label{RDM_and_overM}
\end{eqnarray} 
Therefore, the von Neumann entropy in this bipartite partitioning is given by
\begin{equation}
{\cal S}=-\tr \rho_A \ln \rho_A = -\sum_\tau p_\tau \ln p_\tau,~~~{\rm with}~~~p_\tau=\frac{d^2_\tau}{\sum_\tau d^2_\tau}.
\label{EE_in_terms_of_M}
\end{equation} 
Therefore, we can obtain the von Neumann entropy in terms of the overlap matrix $M$. 
To obtain ${\cal S}$, we need all the eigenvalues of $M$. 

\begin{figure}[tb]
\begin{center}
\vspace{.5cm}
\hspace{-.0cm}\includegraphics[width=0.45\columnwidth]{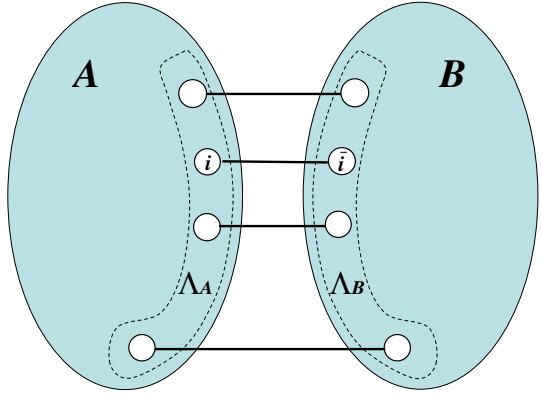}
\vspace{.0cm}
\caption{VBS state on a reflection symmtric graph. $\Lambda_A$ and $\Lambda_B$ denote the sets of sites (vertices) on the boundaries of subgraphs $A$ and $B$. Any site $i \in \Lambda_A$ has a unique reflection symmetric partner ${\bar i}\in B$. }
\label{fig: ref_sym}
\end{center}
\end{figure}

Let us now apply the Schmidt decomposition method to study entanglement in VBS states on graphs with symmetries such as reflection and/or inversion. 
For simplicity, we shall focus on the graphs with reflection symmetry. 
However, one can easily generalize our argument to graphs with inversion symmetry. 
We define reflection symmetry such that any site (vertex) in a subgraph $A$ has a reflection partner in $B$ and vice versa. A graphical example is shown in Fig. \ref{fig: ref_sym}. 
In the Schwinger boson language, the VBS state on a reflection symmetric graph can be written as
\begin{equation}
\fl~~~~~~~~~~
|{\rm VBS} \ra = \prod_{\la i,j\ra\in{\cal B}_A} (a\dag_i b\dag_j-b\dag_i a\dag_j) \prod_{\la i,j\ra\in {\cal B}_B}(a\dag_i b\dag_j-b\dag_i a\dag_j)
\prod_{i \in {\Lambda_A} \atop j \in {\Lambda_B}}(a\dag_i b\dag_j-b\dag_i a\dag_j)|{\rm vac}\ra,
\label{ref_VBS}
\end{equation}
where ${\cal B}_A$ and ${\cal B}_B$ are sets of edges (bonds) in subsystems $A$ and $B$ while $\Lambda_A$ and $\Lambda_B$ are sets of vertices (sites) on the boundaries of $A$ and $B$. 
We note that we only consider the {\it basic } VBS state which is the ground state of the basic AKLT model. In this state, there is one valence bond on any edge $\la i,j \ra$. The generalization of our argument to generic VBS states is straightforward. 
To apply the Schmidt decomposition, we slightly change the above expression by a local gauge transformation: $a_j \to -b_j$ and $b_j \to a_j$ for ${}^\forall j \in B$. 
Then Eq. (\ref{ref_VBS}) becomes
\begin{equation}
\fl~~~~~~~~~~
|{\rm VBS}\ra = \prod_{\la i,j \ra\in{\cal B}_A} (a\dag_i b\dag_j-b\dag_i a\dag_j) \prod_{\la i,j\ra \in {\cal B}_B}(a\dag_i b\dag_j-b\dag_i a\dag_j)
\prod_{i \in {\Lambda_A} \atop j \in {\Lambda_B}}(a\dag_i a\dag_j+b\dag_i b\dag_j)|{\rm vac}\ra.
\end{equation}
Since the graph is reflection symmetric, $i \in \Lambda_A$ necessarily has a reflection symmetric partner ${\bar i}\in \Lambda_B$ (see Fig. \ref{fig: ref_sym}). For simplicity, we henceforth assume that any ${\bar i}$ is  nearest neighbor to $i$ in the graph $G$. 
Therefore, the above expression can be written as follows:
\begin{eqnarray}
\fl 
|\Psi\ra = \prod_{\la i,j\ra \in {\cal B}_A}(a\dag_i b\dag_j-b\dag_i a\dag_j)
\prod_{i \in \Lambda_A} (a\dag_i a\dag_{\bar i}+b\dag_i b\dag_{\bar i})
\prod_{\la i,j \ra\in {\cal B}_B}(a\dag_i b\dag_j-b\dag_i a\dag_j)|{\rm vac}\ra \nonumber \\
\fl
~~~~= \sum_{\{\al \}}\prod_{i \in \Lambda_A}(a\dag_i)^{1/2+\al_i} (b\dag_i)^{1/2-\al_i} (a\dag_{\bar i})^{1/2+\al_i} (b\dag_{\bar i})^{1/2-\al_i}
\prod_{\la i,j \ra\in {\cal B}_A}(a\dag_i b\dag_j-b\dag_i a\dag_j)
\prod_{\la i,j \ra\in {\cal B}_B}(a\dag_i b\dag_j-b\dag_i a\dag_j)|{\rm vac}\ra,
\nonumber \\
\fl
~~~~=\sum_{\{\al \}}|\phi^{[A]}_{\{\al \}} \ra \otimes |\phi^{[B]}_{\{ \al \}}\ra,
\end{eqnarray} 
where $\{\al \}=\{\al_1, \al_2, ..., \al_{|\Lambda_A|} \}$  $(\al_i=\pm 1/2)$ corresponds to the spin state of the boundary state, i.e., degenerate ground state of the {\it block Hamiltonian} defined within the subsystem $A$ or $B$.
The block Hamiltonians are explicitly defined as
\begin{equation}
H_{A} = \sum_{\la k,l \ra \in {\cal B}_A} A(k,l) \pi_{S_k+S_l} (k,l),~~~
H_{B} = \sum_{\la k,l \ra \in {\cal B}_B} A(k,l) \pi_{S_k+S_l} (k,l),
\label{block_Ham}
\end{equation} 
where $\pi_{S_k+S_l} (k,l)$ is defined in Eq. (\ref{projector_pi}).
The states $|\phi^{[A]}_{\{\al \}} \ra$ and $|\phi^{[B]}_{\{ \al \}}\ra$ are given by
\begin{eqnarray}
|\phi^{[A]}_{\{\al \}} \ra &= \prod_{k \in \Lambda_A} (a\dag_k)^{1/2+\al_k} (b\dag_k)^{1/2-\al_k} \prod_{\la i,j\ra \in{\cal B}_A}(a\dag_i b\dag_j-b\dag_i a\dag_j)|{\rm vac}^{[A]}\ra, \nonumber \\
|\phi^{[B]}_{\{ \al \}}\ra &= \prod_{k \in \Lambda_B} (a\dag_k)^{1/2+\al_k} (b\dag_k)^{1/2-\al_k} \prod_{\la i,j \ra\in{\cal B}_B}(a\dag_i b\dag_j-b\dag_i a\dag_j)|{\rm vac}^{[B]}\ra,
\label{phi_A_B}
\end{eqnarray} 
respectively, where $|{\rm vac}^{[A]}\ra$ ($|{\rm vac}^{[B]}\ra$) is the vacuum for bosons in $A$ ($B$) and $|{\rm vac}\ra=|{\rm vac}^{[A]}\ra \otimes |{\rm vac}^{[B]}\ra$. 
Equations in (\ref{phi_A_B}) immediately yield that overlap matrices $M^{[A]}$ and $M^{[B]}$ are the same. The matrix element of $M(=M^{[A]}=M^{[B]})$ is given by
\begin{eqnarray}
M_{\{\al\}, \{\be\}} &= \la {\rm vac}^{[A]}|\prod_{\la i,j \ra\in{\cal B}_A}(a_i b_j-b_i a_j) 
\prod_{k \in \Lambda_A} (a_k)^{1/2+\al_k} (b_k)^{1/2-\al_k}
\nonumber \\
&\times \prod_{k \in \Lambda_A} (a\dag_k)^{1/2+\be_k} (b\dag_k)^{1/2-\be_k}
\prod_{\la i,j \ra\in{\cal B}_A}(a\dag_i b\dag_j-b\dag_i a\dag_j)|{\rm vac}^{[A]}\ra.
\label{integral0}
\end{eqnarray}
The rank of $M$ is $2^{|\Lambda_A|}$, where we denote the number of elements in a set $S$ by $|S|$. The matrix element of $M$ is real, i.e., $(M_{\{\al\}, \{\be\}})^*=M_{\{\al\}, \{\be\}}$, because the commutations of bosons do not produce any complex phases. Therefore, one can easily show the symmetric property $M_{\{\al\}, \{\be\}}=M_{\{\be\}, \{\al\}}$. 

We now rewrite the above expression in terms of the spherical angles. 
This can be done by introducing spin coherent state representation. 
Let us first introduce spinor coordinates $(u_k, v_k)$ at each vertex (site) $k$:
\begin{equation}
u_k=e^{i\phi_k/2} \cos \frac{\theta_k}{2},~~~~~ v_k=e^{-i\phi_k/2} \sin\frac{\theta_k}{2},
\end{equation}
where $0 \le \theta \le \pi$ and $0 \le \phi < 2\pi$. 
Then for a point $\Oh_k \equiv (\sin \theta_k \cos \phi_k, \sin \theta_k \sin \phi_k, \cos \theta_k)$ on the unit sphere, the spin-$S_k$ coherent state is defined as follows:
\begin{equation}
|\Oh_k\ra \equiv \frac{(u_k a\dag_i+v_k b\dag_i)^{2S_k}}{\sqrt{(2S_k)!}}|{\rm vac}^{[k]}\ra.
\end{equation}
Here we have fixed the overall phase (U$(1)$ gauge degree of freedom) since it has no physical content. 
The coherent state is not orthogonal but complete and the resolution of identity is given by 
\begin{equation}
1_{2S_k+1}= \sum^{S_k}_{m=-S_k}|S_k, m \ra \la S_k,m | = \frac{2S_k+1}{4\pi}\int d\Oh_k |\Oh_k\ra \la \Oh_k|,
\end{equation}
where $|S_k, m\ra$ denotes the simultaneous eigenstate of ${\vec S}^2_k$ and $S^z_k$, and $1_{2S_k+1}$ is the $(2S_k+1)$-dimensional identity matrix. 
By inserting this resolution of the identity, Eq. (\ref{integral0}) can be recast in an integral form:
\begin{eqnarray}
M_{\{\al\}, \{\be\}} &= \int \left(\prod_{i \in A} \frac{(2S_i+1)!}{4\pi} d\Oh_i \right)  \prod_{k \in \Lambda_A} u^{1/2+\al_k}_k v^{1/2-\al_k}_k (u^*_k)^{1/2+\be_k} (v^*_k)^{1/2-\be_k} \nonumber \\
&\times \prod_{\la i,j \ra\in{\cal B}_A} \left(\frac{1-\Oh_i \cdot \Oh_j}{2} \right),
\label{integral1}
\end{eqnarray}
where we have used the fact $\la {\rm vac}|a^{S_k-l}_kb^{S_k+l}_k|\Oh_k\ra =\sqrt{(2S_k)!}u^{S-l}_k v^{S+l}_k$. 
Therefore, if we can obtain the matrix element of $M$ by some method such as Monte Carlo integration, the only thing to do is to diagonalize the matrix $M$. Once the eigenvalues of $M$, i.e., $d_\tau$ ($\tau=1, 2, ..., 2^{|\Lambda_A|}$), are obtained, the von Neumann entropy can be calculated through Eq. (\ref{EE_in_terms_of_M}).  
Let us now estimate the upper bound on the von Neumann entropy. 
Suppose that all the eigenvalues of $M$ are the same. This means that the subsystems $A$ and $B$ are {\it maximally entangled}. 
Then, the von Neumann entropy is given by $|\Lambda_A| \ln 2$. This value gives an upper bound on the von Neumann entropy, i.e., ${\cal S} \le |\Lambda_A| \ln 2$. 
However, as we will see in the next section, numerical results indicate 
a deviation from this naive estimate and the von Neumann entropy ${\cal S}$ is strictly less than $|\Lambda_A| \ln 2$ in 2d VBS. 

It is interesting to note that there is a {\it holographic} interpretation of the overlap matrix $M$. The matrix $M$ can be regarded as an operator acting on the auxiliary one dimensional chain which is along the boundary of $A$, i.e., $\Lambda_A$. 
From the fact that $\Oh_k=(\sin \theta_k \cos \phi_k, \sin \theta_k \sin \phi_k, \cos\theta_k)$, one can rewrite $M$ as
\begin{eqnarray}
\fl~~~~~~~~
M = \int \left(\prod_{i \in A} \frac{(2S_i+1)!}{4\pi} d\Oh_i \right)
\prod_{k \in \Lambda_A} \left(\frac{1+\Oh_k \cdot {\vec \sigma}_k}{2} \right)
\prod_{\la i,j \ra\in{\cal B}_A} \left(\frac{1-\Oh_i \cdot \Oh_j}{2} \right).
\label{integral2}
\end{eqnarray}
Here we have used the fact that $M^{\rm T}=M$. Note that the Pauli matrices $\sigma^\alpha_k$ ($\alpha=x, y, z$) act on the auxiliary space, $\{|\uparrow^{[i]}\ra, |\downarrow^{[i]}\ra \}$. 
Then using the following relation:
\begin{equation}
\frac{(S_i+1)(2S_i+1)}{4\pi}\int d\Oh_i\, \Oh_i |\Oh_i\ra \la \Oh_i| =\vec S_i,~~~(a=x, y, z),
\end{equation} 
the overlap matrix can be expressed as a matrix-valued correlation function:
\begin{equation}
M=\la {\rm VBS}^{[A]}|\prod_{i \in \Lambda_A} \left( S_i+\frac{1}{2}+{\vec S}_i \cdot {\vec \sigma}_i \right)|{\rm VBS}^{[A]}\ra,
\label{overlap_and_cor}
\end{equation}
where 
\begin{equation}
|{\rm VBS}^{[A]}\ra \equiv \prod_{\la i,j \ra \in {\cal B}_A}(a\dag_i b\dag_j-b\dag_i a\dag_j )|{\rm vac}^{[A]} \ra
\label{half_VBS}
\end{equation}
Note that the matrix rank of $\vec S_i$ for $i \in \Lambda_A$ is $2S_i-1$ while that for $i \in A\setminus \Lambda_A$ is $2S_i+1$. 
One can interpret the overlap matrix as a Hamiltonian acting on $\sigma$-spin 
space. The matrix element of this Hamiltonian is governed by the boundary correlation in the half-VBS state given in Eq. (\ref{half_VBS}). The naive upper bound value of the von Neumann entropy, $|\Lambda_A| \ln 2$, corresponds to the strongly disordered limit where any boundary correlation among $S$-spins vanishes. However, as shown numerically in \cite{Niggemann1, Niggemann2}, the spin-spin correlation functions at small distances are nonzero in the VBS-type state even though they decay exponentially fast. Therefore, the leading term in
Eq. (\ref{overlap_and_cor}) is given by a constant plus a term proportional to the spin-$\frac{1}{2}$ Heisenberg Hamiltonian whose coupling $J$ is given by the nearest neighbor $S$-spin correlation function. The eigenvalues of this matrix can be different and hence the entanglement spectrum may not be flat. 
Therefore, it is plausible that the von Neumann entropy is strictly less than $|\Lambda_A|\ln 2$ even in the 2d infinite system. To support this conjecture, we perform numerical and analytical calculations for the VBS states on square and hexagonal lattices in the following sections. 

\section{Numerical analysis of 2d VBS states}

In the previous section, we derived the integral formula for the overlap matrix (Eq. (\ref{integral1})) which helps us to reduce computational cost. 
Let us explain this point in more detail. Suppose that the graph $G$ has $|V|$ vertices and every vertex has the same spin $S$. Then, the needed dimension for representing the VBS state on $G$ 
is $(2S+1)^{|V|}$ without consideration of symmetry. 
However, if we focus on the von Neumann entropy of reflection symmetric VBS state, the needed dimension is greatly reduced from $(2S+1)^{|V|}$ to  $2^{|\Lambda_A|}$, where $|\Lambda_A|$ denotes the number of sites on the boundary between $A$ and $B$. 
Therefore, we can study the von Neumann entropy of 2d VBS state on a relatively large system by combining Monte Carlo integration and exact diagonalization. After obtaining the overlap matrix and all its eigenvalues, the von Neumann entropy can be calculated according to Eq. (\ref{EE_in_terms_of_M}).
In this section, we calculate von Neumann entropies of VBS states on square and hexagonal lattices with open boundary conditions using Monte Carlo integration.

\subsection{von Neumann entropy of VBS state on square lattice}

\begin{figure}[tb]
\begin{center}
\vspace{.5cm}
\hspace{-.0cm}\includegraphics[width=0.6\columnwidth]{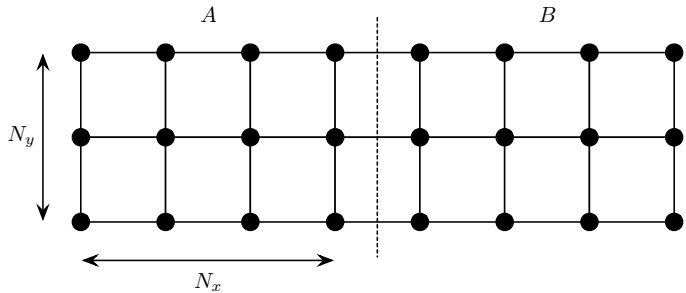}
\vspace{.0cm}
\caption{Square lattice with open boundary conditions. $N_{\rm x}$ and $N_{\rm y}$ denote the numbers of sites along $x$- and $y$-axes in $A$, respectively.}
\label{fig:squarelattice}
\end{center}
\end{figure}

\begin{figure}[tb]
\begin{center}
\vspace{.5cm}
\hspace{-.0cm}\includegraphics[width=0.6\columnwidth]{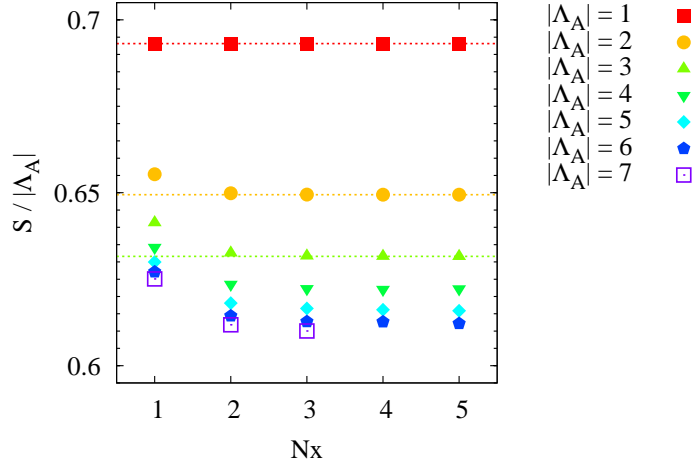}
\vspace{.0cm}
\caption{The von Neumann entropy per valence bond as a function of $N_{\rm x}$ for the square VBS states with various $N_y$. The dotted lines indicate the exact results in $N_x \to \infty$ limit.}
\label{graph:EE-square-nx}
\end{center}
\end{figure}

\begin{figure}[tb]
 \begin{center}
  \vspace{.5cm}
  \hspace{-.0cm}
  \includegraphics[width=\columnwidth]{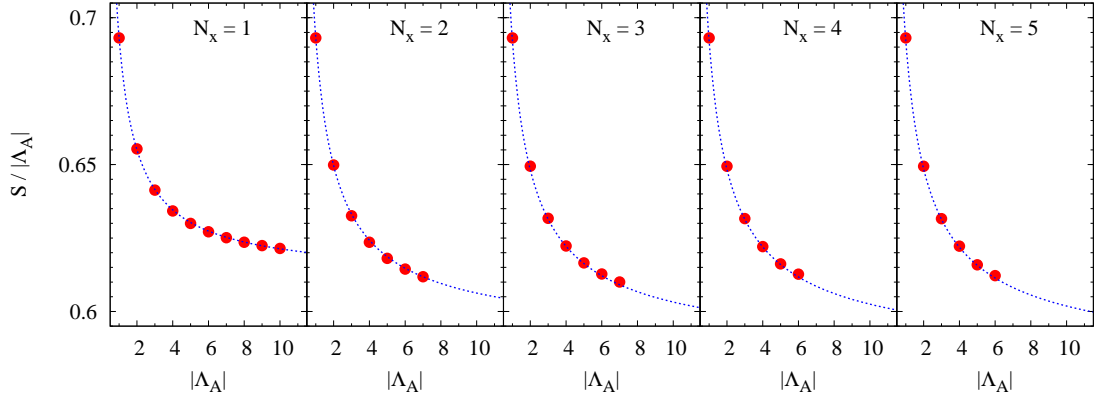}
  \vspace{.0cm}
  \caption{Entanglement entropy per valence bond as a function of
  $N_{\rm y}$ for the square lattice VBS states with various $N_x$. Blue curves represent the fitting curves (Eq. (\ref{eq:fitting})).}
  \label{graph:EE-square-ny}
 \end{center}
\end{figure}

We first study the von Neumann entropy of the VBS state on square lattices with open boundary conditions. 
The reflection axis of the square lattice is taken to be the boundary between $A$ and $B$ as shown in Fig. \ref{fig:squarelattice}.
$N_x$ in Fig. \ref{fig:squarelattice} is the number of sites along $x$-axis while $N_y$ is that along $y$-axis. 
For square lattice, the number of boundary sites $|\Lambda_A|$ is the same as $N_y$, i.e., $|\Lambda_A|=N_y$. 
We first study the $N_x$-dependence of the von Neumann entropy.
Figure \ref{graph:EE-square-nx} shows the von Neumann entropy per boundary site, i.e., valence bond, for various $N_y$ as a function of $N_x$. 
The von Neumann entropy per valence bond, ${\cal S}/|\Lambda_A|$, rapidly converges to certain values depending on $N_y$. 
The case of $N_y=1$ reduces to the linear AKLT chain along $x$-axis. 
In this case, ${\cal S}/|\Lambda_A|={\cal S}$ monotonically decreases with increasing $N_x$ and approaches to $\ln 2~(=0.6931472...)$ exponentially fast. This correctly reproduces the exact result in \cite{Fan_Korepin1}. 
The obtained results show that ${\cal S}/|\Lambda_A|$ for $N_y = 2$ and $N_y = 3$ approach to $0.6494348$ and $0.6315983$, respectively.
These values are consistent with the analytical results shown in the next section. 
Next we consider the $N_y$-dependence of the von Neumann entropy.
Figure \ref{graph:EE-square-ny} shows the von Neumann entropy per boundary site as a function of $N_y$. 
We find that ${\cal S}/|\Lambda_A|$ is well fitted by the following function:
\begin{eqnarray}
 \label{eq:fitting}
 \frac{{\cal S}}{|\Lambda_A|} = \frac{C}{|\Lambda_A|^\Delta} + \alpha,
\end{eqnarray}
where 
$C$, $\Delta$ and $\alpha$ are fitting parameters. 
The constant term $\alpha$ means the von Neumann entropy per valence bond in the limit of $|\Lambda_A| \to \infty$. Note that $C$, $\Delta$ and $\alpha$ depend on $N_x$.
The blue curves in Fig. \ref{graph:EE-square-ny} represent the fitting curves. 
The obtained numerical data show good agreement with the power law behavior assumed in Eq. (\ref{eq:fitting}). 
%
\begin{table}
\begin{center}
\begin{tabular}{|c|c|c|c|}
 \hline
 & $C$ & $\Delta$ & $\alpha$ \\
 \hline
 $N_x = 1$ & $0.0821(4)$ & $0.90(1)$ & $0.6110(4)$ \\
 \hline
 $N_x = 2$ & $0.104(1)$ & $0.78(2)$ & $0.589(1)$ \\
 \hline
 $N_x = 3$ & $0.108(1)$ & $0.75(2)$ & $0.584(1)$ \\
 \hline 
 $N_x = 4$ & $0.110(2)$ & $0.73(3)$ & $0.582(2)$ \\
 \hline
 $N_x = 5$ & $0.112(1)$ & $0.71(2)$ & $0.580(1)$ \\
 \hline
\end{tabular} 
\end{center}
 \caption{The coefficients of the fitting function Eq. (\ref{eq:fitting}) for square lattices.}
 \label{table:EE-square}
\end{table}
Table \ref{table:EE-square} shows the fitting parameters $C$, $\Delta$ and $\alpha$ for $N_x = 1, 2, 3, 4$ and $5$.
As $N_x$ increases, the coefficient $C$ increases and $\Delta$ decreases.
The non-universal coefficient for the area law, $\alpha$, decreases very slowly when $N_x$ increases. Suppose that $\Delta$ does not vanish in the large $N_x$ limit. Then, we can rewrite Eq. (\ref{eq:fitting}) as
\begin{equation}
{\cal S}=\alpha |\Lambda_A| + C|\Lambda_A|^{1-\Delta}.
\end{equation}
The first term simply means the area law while the second term is a tiny correction from it. Our numerical results indicate the coefficient $\alpha$ is strictly less than $\ln 2$ in the $N_x \to \infty$ limit.

\subsection{von Neumann entropy of VBS state on hexagonal lattice}
\label{sec: MC hexagonal}

\begin{figure}[tb]
\begin{center}
\vspace{.5cm}
\hspace{-.0cm}\includegraphics[width=0.6\columnwidth]{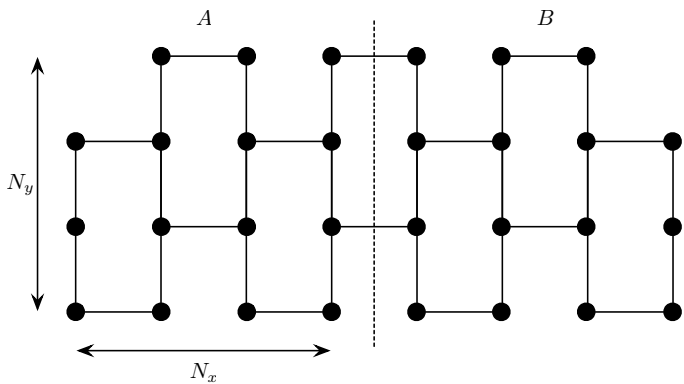}
\vspace{.0cm}
\caption{Hexagonal lattice with open boundary condition. $N_x$ and $N_y$ denote the numbers of the sites along $x$- and $y$- axes in $A$, respectively.}
\label{fig:hexlattice}
\end{center}
\end{figure}

\begin{figure}[tb]
\begin{center}
\vspace{.5cm}
\hspace{-.0cm}\includegraphics[width=0.6\columnwidth]{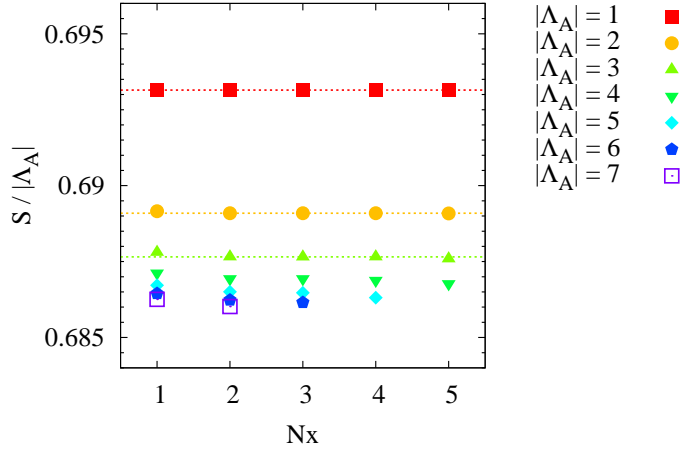}
\vspace{.0cm}
\caption{Entanglement entropy per valence bond as a function of $N_x$ for the hexagonal VBS states with various $|\Lambda_A|$. The dotted lines indicate the exact results in $N_x \to \infty$ limit.}
\label{graph:EE-hex-nx}
\end{center}
\end{figure}

\begin{figure}[tb]
 \begin{center}
  \vspace{.5cm}
  \hspace{-.0cm}
  \includegraphics[width=\columnwidth]{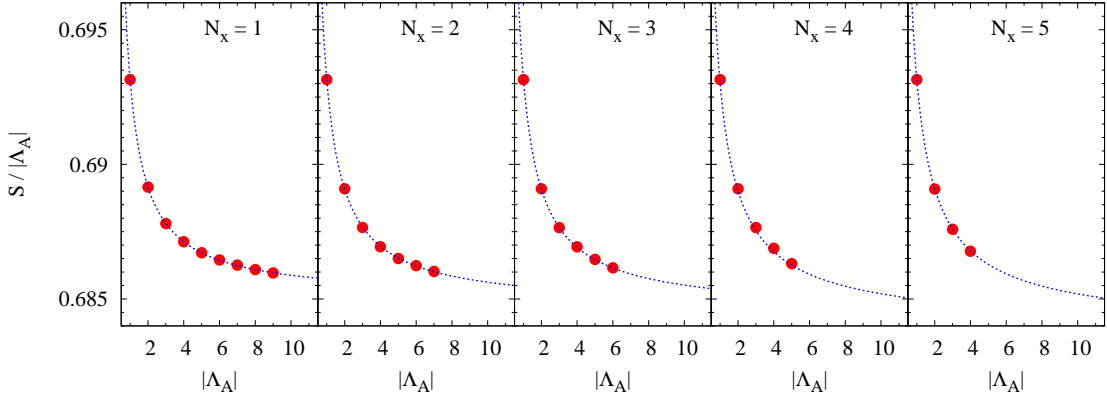}
  \vspace{.0cm}
  \caption{The von Neumann entropy per valence bond as a function of
  $N_y$ for the hexagonal VBS states with various $N_x$. Blue curves represent the fitting curves.}
  \label{graph:EE-hex-ny}
 \end{center}
\end{figure}

In this subsection, we consider the von Neumann entropy of the VBS state on hexagonal lattices with open boundary conditions. 
Similarly to the case of the square lattice, subsystems $A$ and $B$ are partitioned by the reflection axis shown in Fig. \ref{fig:hexlattice}.
$N_x$ in Fig. \ref{fig:hexlattice} is the number of sites along $x$-axis while $N_y$ is that along $y$-axis. 
Since we focus on even-$N_y$ cases in the present study, the number of the boundary sites is $|\Lambda_A|= N_y/2$.
%
We first study the $N_x$-dependence of von Neumann entropy. 
The results of the von Neumann entropy per boundary site, i.e., valence bond, 
are shown in Fig. \ref{graph:EE-hex-nx}. 
The case of $|\Lambda_A|=1$ reduces to the $S=1$ AKLT chain and reproduces the exact result in~\cite{Fan_Korepin1}.  
The von Neumann entropy per valence bond, ${\cal S}/|\Lambda_A|$, for $N_y = 2$ and $3$ approach to $0.6890926$ and $0.6876522$, respectively.
These values also consistent with the analytical results shown in Section \ref{sec: alg}.
Next we study the $N_y$-dependence of the von Neumann entropy. 
The obtained results are shown in Fig. \ref{graph:EE-hex-ny}. 
Those results are again well fitted by the scaling function given by Eq. (\ref{eq:fitting}). Those fitting curves are shown by the blue curves in Fig. \ref{graph:EE-hex-ny}. 
The fitting parameters $C$, $\Delta$, and $\alpha$ for $N_x=1,2,3,4$, and $5$ are summarized in 
Table \ref{table:EE-hexagonal}. 
As $N_x$ increases, $C$ increases and $\Delta$ and $\alpha$ decreases. 
Assuming $\Delta$ is nonvanishing in the large $N_x$ limit, ${\cal S}/|\Lambda_A|$ is strictly less than $\ln 2$ even in the infinite 2d system. 
Therefore, our numerical results again supports the conjecture given in Sec. \ref{sec: schmidt}. 
\begin{table}[h]
\begin{center}
\begin{tabular}{|c|c|c|c|}
 \hline
 & $C$ & $\Delta$ & $\alpha$ \\
 \hline
 $N_x = 1$ & $0.00812(1)$ & $0.974(4)$ & $0.68502(1)$ \\
 \hline
 $N_x = 2$ & $0.00849(4)$ & $0.94(1)$ & $0.68465(5)$ \\
 \hline
 $N_x = 3$ & $0.00870(3)$ & $0.904(7)$ & $0.68443(3)$ \\
 \hline
 $N_x = 4$ & $0.0092(2)$ & $0.82(3)$ & $0.6838(2)$ \\
 \hline
 $N_x = 5$ & $0.00941(7)$ & $0.81(1)$ & $0.68373(7)$ \\
 \hline
\end{tabular}
\end{center}
 \caption{The coefficients of the fitting function Eq. (\ref{eq:fitting}) for hexagonal lattices.}
 \label{table:EE-hexagonal}
\end{table}
%
\section{Algebraic analysis of VBS ladders}
\label{sec: alg}
In this section, we shall consider the VBS states on various ladders and study the von Neumann entropy via another way. 
It is convenient to introduce the following function of $\{ \vs_i \}$:
\begin{equation}
Z(\vs_1, ..., \vs_{|\Lambda_A|}) 
=\int \left( \prod_{i \in A} \frac{d\Oh_i}{4\pi}\right)
\prod_{k \in \Lambda_A} (1+\Oh_k \cdot \vs_k) 
\prod_{\la i,j \ra \in {\cal B}_A} (1-\Oh_i \cdot \Oh_j),
\label{eq: def of Z}
\end{equation}
where $\vs_i$'s are classical or operator-valued vectors depending on the context. The overlap matrix given in Eq. (\ref{integral2}) is proportional to $Z(\vec \sigma_1, ..., \vec \sigma_{|\Lambda_A|})$. 
Since the only difference between $Z$ and $M$ is the overall factor, the eigenvalues of the reduced density matrix equal to those of the following matrix:
\begin{equation}
{\tilde \rho}_A = \frac{Z(\vec \sigma_1, ..., \vec \sigma_{|\Lambda_A|})}{{\rm tr} Z(\vec \sigma_1, ..., \vec \sigma_{|\Lambda_A|})},
\label{eq: rho_tilde}
\end{equation}
where the trace is taken over the $\sigma$-spin spaces. The von Neumann entropy is expressed in terms of $\tilde \rho_A$ as
\begin{equation}
{\cal S}=-{\rm tr}{\tilde \rho}_A \ln {\tilde \rho}_A.
\end{equation}
Therefore, we shall henceforth use the matrix $Z$ instead of $M$.
Our strategy is to find a relation between $Z$ matrices for different size systems. Once we find the relation, we can construct $Z$ matrices recursively and obtain the von Neumann entropy exactly. 

\subsection{VBS states on square ladders}
\label{sec: square ladders}
\subsubsection{2-leg square ladder}
\begin{figure}[tb]
\begin{center}
\vspace{.5cm}
\hspace{-.0cm}\includegraphics[width=0.55\columnwidth]{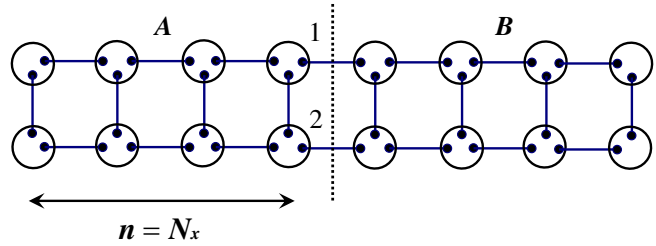}
\vspace{.0cm}
\caption{VBS state on a 2-leg ladder. It is cut by the reflection line indicated by the broken line.}
\label{fig: VBS_ladder}
\end{center}
\end{figure}

Let us first study the VBS state on the ladder shown in Fig. \ref{fig: VBS_ladder}. 
In this case, the overlap matrix is proportional to
\begin{equation}
\fl ~~~~~~~~
Z_{n}(\vec \sigma_1, \vec \sigma_2)=\int \left( \prod_{i \in A} \frac{d\Oh_i}{4\pi}\right)
(1+\Oh_1 \cdot \vec \sigma_1) (1+\Oh_2 \cdot \vec \sigma_2)
\prod_{\la i,j \ra \in {\cal B}_A} (1-\Oh_i \cdot \Oh_j),
\label{eq: integral=2leg_square}
\end{equation}
where $n=N_x$ denotes the horizontal length of $A$ and boundary sites 1 and 2 are shown in Fig. \ref{fig:squarelattice}. The product in the above equation can be expanded as
\begin{equation}
\prod_{\la i,j \ra\in {\cal B}_A} (1-\Oh_i \cdot \Oh_j)=\sum_{\Gamma \subset {\cal B}_A} \prod_{\la i,j \ra\in \Gamma} (-\Oh_i \cdot \Oh_j),
\end{equation}
where the sum runs over all subsets $\Gamma$ of ${\cal B}_A$.
Since the measure of integration is invariant under the local change of variable, $\Oh_i \to -\Oh_i$, we see that the integrand in Eq. (\ref{eq: integral=2leg_square}) must contain even number of $\Oh_i$'s. 
(We give a graphical and combinatorial explanation of this fact in terms of self-avoiding loops and strands in \ref{sec:loop model}.)
Therefore, $Z_n (\vs_1, \vs_2)$ must be the following form:
\begin{equation}
Z_n (\vs_1, \vs_2) = {\ssf a}_n - {\ssf b}_n \vs_1 \cdot \vs_2.
\label{eq: Zmatrix_2leg_square}
\end{equation}
Then the $4 \times 4$ matrix $Z_n(\vec \sigma_1, \vec \sigma_2)$ is written in terms ${\sf a}_n$ and ${\sf b}_n$ as
\begin{equation}
\fl ~~~~~~~~
Z_n (\vec \sigma_1, \vec \sigma_2)=
\left( \begin{array}{cccc}
{\ssf a}_n-{\ssf b}_n & 0 & 0 & 0 \\
0 & {\ssf a}_n+{\ssf b}_n & -2{\ssf b}_n & 0 \\
0 & -2{\ssf b}_n & {\ssf a}_n+{\ssf b}_n & 0 \\
0 & 0 & 0 & {\ssf a}_n-{\ssf b}_n
\end{array} \right),
\end{equation}
where ${\ssf a}_n$ and ${\ssf b}_n$ are coefficients depending on $n$. 
The eigenvalues of the matrix $Z_n(\vec \sigma_1, \vec \sigma_2)$ are ${\ssf a}_n+3{\ssf b}_n$ and 3-fold degenerate ${\ssf a}_n-{\ssf b}_n$.
We shall now compute $Z_n$ by induction on $n$. 
The relation between $Z_{n+1}$ and $Z_n$ is given by
\begin{eqnarray}
\fl 
Z_{n+1}(\vec \sigma_1, \vec \sigma_2) &=
\int \frac{d\Oh_1}{4\pi} \frac{d\Oh_2}{4\pi}
(1+\Oh_1 \cdot \vec \sigma_1) (1+\Oh_2 \cdot \vec \sigma_2) (1-\Oh_1 \cdot \Oh_2)\, Z_n (\Oh_1, \Oh_2) \nonumber \\
&= \int \frac{d\Oh_1}{4\pi} \frac{d\Oh_2}{4\pi}
(1+\Oh_1 \cdot \vec \sigma_1) (1+\Oh_2 \cdot \vec \sigma_2) (1-\Oh_1 \cdot \Oh_2) ({\ssf a}_n-{\ssf b}_n \Oh_1 \cdot \Oh_2) \nonumber \\
&= \int \frac{d\Oh_1}{4\pi} \frac{d\Oh_2}{4\pi}
\left[{\ssf a}_n+{\ssf b}_n (\Oh_1 \cdot \Oh_2)^2 
-({\ssf a}_n+{\ssf b}_n)(\Oh_1 \cdot \vec \sigma_1)(\Oh_2 \cdot \vec \sigma_2)(\Oh_1 \cdot \Oh_2) \right]. \nonumber \\
\end{eqnarray}
Here we have used the fact that the integration over odd number of $\Oh_i$'s is zero (see \ref{sec:loop model}). 
To perform the integration over $\Oh_1$ and $\Oh_2$, the following relation shown in the lemma 3.3 in Ref.~\cite{Kennedy_Lieb_Tasaki} is useful: 
\begin{equation}
\int \frac{d\Oh_i}{4\pi} (\vs_1 \cdot \Oh_i) (\Oh_i \cdot \vs_2)=q \vs_1 \cdot \vs_2,
\label{3.3}
\end{equation}
where $q=1/3$. Using this formula, we obtain 
\begin{equation}
Z_{n+1}(\vec \sigma_1, \vec \sigma_2)\equiv 
{\ssf a}_{n+1}-{\ssf b}_{n+1}\vec \sigma_1 \cdot \vec \sigma_2 = ({\ssf a}_n+q {\ssf b}_n)-q^2({\ssf a}_n+{\ssf b}_n)\vec \sigma_1 \cdot \vec \sigma_2.
\end{equation}
Therefore, the recursion relation between $\{ {\ssf a}_{n+1}, {\ssf b}_{n+1} \}$ and $\{ {\ssf a}_n, {\ssf b}_n \}$ is given by
\begin{equation}
\left( \begin{array}{c}
{\ssf a}_{n+1} \\ {\ssf b}_{n+1}
\end{array}\right)=\left( \begin{array}{cc}
1 & q \\
q^2 & q^2
\end{array}\right)\left( \begin{array}{c}
{\ssf a}_{n} \\ {\ssf b}_{n}
\end{array}\right).
\label{rec_2leg_square}
\end{equation}
Here, the initial values ${\ssf a}_0$ and ${\ssf b}_0$ are $1$ and $0$, respectively. 
Equation (\ref{rec_2leg_square}) can be easily solved and ${\ssf a}_n$ and ${\ssf b}_n$ for arbitrary $n$ are given by
\begin{eqnarray}
{\ssf a}_n = \frac{1}{2\sqrt{19}}  [4(z_+^{n}-z_-^{n})+\sqrt{19}(z_+^{n}+z_-^{n})],~~~~
{\ssf b}_n = \frac{1}{2 \sqrt{19}} (z_+^{n}-z_-^{n}),
\end{eqnarray}
with $z_{\pm} =(5 \pm \sqrt{19})/9 $. 
According to Eq. (\ref{eq: rho_tilde}), the eigenvalues of the reduced density matrix are given by
\begin{eqnarray} 
p_1(n) = 
\frac{(1+3x_n)^2}{(1+3x_n)^2+3(1-x_n)^2}, \\
p_2(n) = p_3(n) = p_4(n) = 
\frac{(1-x_n)^2}{(1+3x_n)^2+3(1-x_n)^2},
\end{eqnarray}
where $x_n \equiv {\ssf b}_n/{\ssf a}_n$. From the above explicit expressions, one can calculate the von Neumann entropy for any $n$. The exact results are totally consistent with those obtained from the numerical approach in the previous section (see Table \ref{tab: square ladders}). 
In the large-block-size limit, $x_n \to 1/(4+\sqrt{19})$.
In this limit, the von Neumann entropy is obtained as 
\begin{equation}
{\cal S}^{\rm 2leg}_{\infty}=-p_1(\infty) \ln p_1 (\infty) -3 p_2(\infty) \ln p_2(\infty)=1.2988696.
\end{equation}
Therefore, the von Neumann entropy per valence bond is given by ${\cal S}^{\rm 2leg}_{\infty}/2=0.6494348$, which is strictly less than $\ln 2=0.6931471...$ even in the limit of $N_x \to \infty$.
\begin{table}[h]
\begin{center}
\vspace{0.1in}
\begin{tabular}{c|c|c|c|c|c} \hline
   & $N_x=1$ & $N_x=2$ & $N_x=3$ & $N_x=4$ & $N_x=5$ \\ 
\hline
$\shelf{N_y=2}{\rm Exact}{\rm MC}$ & $\vectorize{0.6553433}{0.6553431}$ & $\vectorize{0.6498531}{0.6498533}$ & $\vectorize{0.6494635}{0.6494621}$ & $\vectorize{0.6494368}{0.6494342}$ & $\vectorize{0.6494349}{0.6494373}$ \\
\hline
$\shelf{N_y=3}{\rm Exact}{\rm MC}$ & $\vectorize{0.6413153}{0.6413145}$ & $\vectorize{0.6325619}{0.6325626}$ & 
$\vectorize{0.6316999}{0.6316999}$& $\vectorize{0.6316095}{0.6316080}$ & $\vectorize{0.6315995}{0.6315866}$\\
\hline 
\end{tabular}
\caption{${\cal S}/|\Lambda_A|$ with $|\Lambda_A|=N_y$ for various $N_x$ obtained by algebraic (Exact) and numerical (MC) methods. The numerical values are rounded off to 8 decimal places.} 
\label{tab: square ladders} 
\end{center}
\end{table}
\subsubsection{3-leg square ladder}
\label{sec: 3-leg square ladder}
\begin{figure}[htb]
\begin{center}
\vspace{.5cm}
\hspace{-.0cm}\includegraphics[width=0.55\columnwidth]{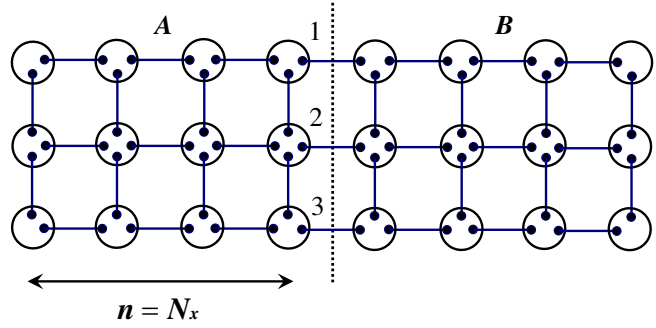}
\vspace{.0cm}
\caption{VBS state on a 3-leg square ladder.}
\label{fig: 3leg_ladder}
\end{center}
\end{figure}

Let us next study the VBS state on the 3-leg ladder shown in Fig. \ref{fig: 3leg_ladder}. 
In this case, the overlap matrix is proportional to
\begin{eqnarray}
\fl 
Z_{n}(\vec \sigma_1, \vec \sigma_2, \vec \sigma_3)=\int \left( \prod_{i \in A} \frac{d\Oh_i}{4\pi}\right) (1+\Oh_1 \cdot \vec \sigma_1) (1+\Oh_2 \cdot \vec \sigma_2)(1+\Oh_3 \cdot \vec \sigma_3)
\prod_{\la i,j \ra \in {\cal B}_A} (1-\Oh_i \cdot \Oh_j), \nonumber
\end{eqnarray}
where the boundary sites 1, 2, and 3 are shown in Fig. \ref{fig: 3leg_ladder}.
Let us set 
\begin{equation}
Z_n(\vs_1, \vs_2, \vs_3)
={\ssf a}_n-{\ssf b}_n \vs_1 \cdot \vs_2 
-{\ssf c}_n \vs_2 \cdot \vs_3 +{\ssf d}_n \vs_1 \cdot \vs_3.
\end{equation}
Note that ${\ssf a}_0=1$, ${\ssf b}_0={\ssf c}_0={\ssf d}_0=0$ and 
\begin{equation}
Z_n(\vec \sigma_1, \vec \sigma_2, \vec \sigma_3)=
{\ssf a}_n-{\ssf b}_n \vec \sigma_1 \cdot \vec \sigma_2 
-{\ssf c}_n \vec \sigma_2 \cdot \vec \sigma_3 +{\ssf d}_n \vec \sigma_1 \cdot \vec \sigma_3
\label{Z_n_3leg}
\end{equation}
is $8\times 8$ matrix. 
We shall now compute $Z_n$ by induction on $n$. The relation between $Z_{n+1}$ and $Z_n$ is given by
\begin{eqnarray}
\fl 
Z_{n+1}(\vec \sigma_1, \vec \sigma_2, \vec \sigma_3)
=& \int \left( \prod^3_{i=1}\frac{d\Oh_i}{4\pi} \right)
(1+\Oh_1 \cdot \vec \sigma_1) (1+\Oh_2 \cdot \vec \sigma_2)(1+\Oh_3 \cdot \vec \sigma_3) \nonumber \\
&\times (1-\Oh_1 \cdot \Oh_2)(1-\Oh_2 \cdot \Oh_3)
Z_{n  }(\Oh_1, \Oh_2, \Oh_3).
\end{eqnarray}
Now we will use Eq. (\ref{3.3}). At the beginning, we exclude from the formula above the vector $\Oh_1$, then $\Oh_3$, and finally $\Oh_2$. As a result, we will have the recursion relation between  $\{{\ssf a}_{n+1}, {\ssf b}_{n+1}, {\ssf c}_{n+1}, {\ssf d}_{n+1} \}$ and $\{{\ssf a}_{n}, {\ssf b}_{n}, {\ssf c}_{n}, {\ssf d}_{n} \}$ as
\begin{equation}
\left(\begin{array}{c} {\ssf a}_{n+1} \\ {\ssf b}_{n+1} \\ {\ssf c}_{n+1} \\ {\ssf d}_{n+1} \end{array}\right)=
\left(\begin{array}{cccc} 
1 & q & q & q^2 \\
q^2 & q^2 & q^3 & q^3 \\
q^2 & q^2 & q^3 & q^3 \\
q^3 & q^3 & q^3 & q^2
\end{array}\right)
\left(\begin{array}{c} {\ssf a}_{n} \\ {\ssf b}_{n} \\ {\ssf c}_{n} \\ {\ssf d}_{n} \end{array}\right),
\end{equation}
where $q=1/3$. Recalling that ${\ssf a}_0=1$, ${\ssf b}_0={\ssf c}_0={\ssf d}_0=0$, we immediately note that ${\ssf b}_n={\ssf c}_n$ for all $n \ge 0$. Therefore, one obtains
\begin{equation}
\left(\begin{array}{c} {\ssf a}_{n+1} \\ {\ssf b}_{n+1} \\ {\ssf d}_{n+1} \end{array}\right)=
\left(\begin{array}{cccc} 
1 & 2q & q^2 \\
q^2 & q^2 + q^3 & q^3 \\
q^2 & 2q^3 & q^2 
\end{array}\right)
\left(\begin{array}{c} {\ssf a}_{n} \\ {\ssf b}_{n} \\ {\ssf d}_{n} \end{array}\right). 
\label{rec_3leg}
\end{equation}
It is now straightforward to obtain the coefficients ${\ssf a}_n$, ${\ssf b}_n$, and ${\ssf d}_n$ by using the above recursion relation.

We shall now calculate the eigenvalues of the reduced density matrix and obtain the von Neumann entropies. One can confirm that the eigenvalues of the $8 \times 8$ matrix $Z_n$ in Eq. (\ref{Z_n_3leg}) are given by
\begin{eqnarray}
\lambda_{1,n}=\lambda_{2,n} ={\ssf a}_n-3{\ssf d}_n, \\
\lambda_{3,n}=\lambda_{4,n}=\lambda_{5,n}=\lambda_{6,n} ={\ssf a}_n-2{\ssf b}_n+{\ssf d}_n, \\
\lambda_{7,n}=\lambda_{8,n} ={\ssf a}_n+4{\ssf b}_n+{\ssf d}_n. 
\end{eqnarray}
Therefore, the eigenvalues of the reduced density matrix are given by
\begin{eqnarray}
\fl 
p_1(n)=p_2(n)= \frac{(1-3y_n)^2}{2(1-3y_n)^2+4(1-2x_n+y_n)^2+2(1+4x_n+y_n)^2}, \\
\fl
p_3(n)=\cdots =p_6(n)=\frac{(1-2x_n+y_n)^2}{2(1-3y_n)^2+4(1-2x_n+y_n)^2+2(1+4x_n+y_n)^2},\\
\fl
p_7(n)=p_8(n)=\frac{(1+4x_n+y_n)^2}{2(1-3y_n)^2+4(1-2x_n+y_n)^2+2(1+4x_n+y_n)^2},
\end{eqnarray}
where 
\bege
x_n=\frac{{\ssf b}_n}{{\ssf a}_n},~~~ y_n=\frac{{\ssf d}_n}{{\ssf a}_n}.
\ende
Then the von Neumann entropy is expressed as 
\begin{equation}
{\cal S}^{\rm 3leg}_n = -\sum^8_{i=1} p_i(n) \ln p_i(n).  
\end{equation} 
The results are summarized in Table \ref{tab: square ladders}. 
One can again see that the results obtained by Monte Carlo integration are consistent with the exact results. Finally, let us consider the large-block-size limit. 
Since the matrix appearing in the recursion relation (\ref{rec_3leg}) is a  positive and irreducible matrix, the largest eigenvalue is unique. 
Let $\vec v_{\rm PF}=(v_1, v_2, v_3)^{\rm T}$ be the corresponding Perron-Frobenius vector. Note that ${}^{\rm T}$ denotes matrix transpose. 
Then $x_n$ and $y_n$ in the limit of $n \to \infty$ are given by
\begin{equation}
x \equiv \lim_{n \to \infty} x_n = \frac{v_2}{v_1},~~~~~ y \equiv \lim_{n \to \infty} y_n= \frac{v_3}{v_1}.
\end{equation}
From the numerically obtained $v_{\rm PF}$, numerical values for $x$ and $y$ are given by
\begin{equation}
x=0.1203998879...,~~~~~y=0.0471631199....
\end{equation}
and hence the von Neumann entropy in the large-block-size limit is obtained as
\begin{equation}
{\cal S}^{\rm 3leg}_\infty = -\sum^8_{i=1} p_i(\infty) \ln p_i(\infty)=  1.8947948. 
\end{equation} 
Therefore, the von Neumann entropy per valence bond is ${\cal S}^{\rm 2leg}_\infty/3=0.6315983$, which is again less than $\ln 2$.

\subsection{VBS states on hexagonal ladders}
\subsubsection{2-leg hexagonal ladder}
\begin{figure}[htb]
\begin{center}
\vspace{.5cm}
\hspace{-.0cm}\includegraphics[width=0.6\columnwidth]{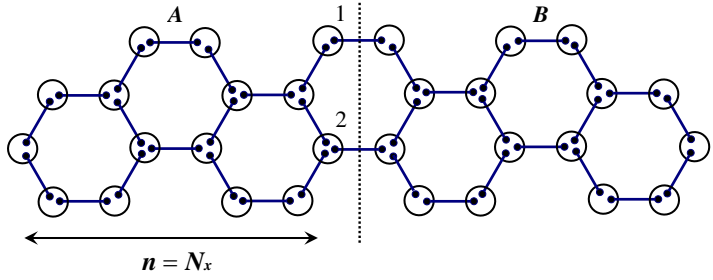}
\vspace{.0cm}
\caption{VBS state on a 2-leg hexagonal ladder. It is cut by the reflection line indicated by the broken line.}
\label{fig: hexagonal_VBS_ladder}
\end{center}
\end{figure}

Let us now study the VBS state on the hexagonal lattice shown in Fig. \ref{fig: hexagonal_VBS_ladder}, which corresponds to $N_y=4$. 
\\
Along the same lines as the square-ladder case, the $Z$ matrix is written as
\begin{equation}
Z_n(\vec \sigma_1, \vec \sigma_2)= {\ssf a}_n + {\ssf b}_n \vec \sigma_1 \cdot \vec \sigma_2,
\label{hex_Zn_2leg}
\end{equation}
and the recursion relation between $\{{\ssf a}_{n+1}, {\ssf b}_{n+1} \}$ and $\{ {\ssf a}_n, {\ssf b}_n \}$ is given by
\begin{equation}
\left( \begin{array}{c} {\ssf a}_{n+1} \\ {\ssf b}_{n+1} \end{array}\right)=
\left( \begin{array}{cc} 1 & q^2 \\ q^3 & q^4 \end{array}\right)
\left( \begin{array}{c} {\ssf a}_{n} \\ {\ssf b}_{n} \end{array}\right),
\end{equation}
where $q=1/3$ and the initial values ${\ssf a}_0$ and ${\ssf b}_0$ are $1$ and $0$, respectively. The above recursion relation can be solved and the coefficients ${\ssf a}_n$ and ${\ssf b}_n$ for arbitrary $n$ is given by
\begin{equation}
{\ssf a}_n = \frac{1}{2\sqrt{1627}}[40 (z_+^n-z_-^n) +\sqrt{1627}(z_+^n+z_-^n)],~~~~
{\ssf b}_n = \frac{3}{2\sqrt{1627}} (z_+^n-z_-^n)
\end{equation}
with 
$z_{\pm}= (41 \pm \sqrt{1627})/81$. 
Since the eigenvalues of $Z_n$ in Eq. (\ref{hex_Zn_2leg}) are ${\ssf a}_n-3{\ssf b}_n$ and 3-fold degenerate ${\ssf a}_n+{\ssf b}_n$, the eigenvalues of the reduced density matrix is given by
\begin{eqnarray}
p_1(n) &=& \frac{(1-3x_n)^2}{(1-3x_n)^2+3(1+x_n)^2}, \\
p_2(n) &=& p_3(n)=p_4(n)= \frac{(1+3x_n)^2}{(1-3x_n)^2+3(1+x_n)^2},
\end{eqnarray}
where $x_n={\ssf b}_n/{\ssf a}_n$. 
The results of the von Neumann entropy are summarized in Table \ref{tab: hexagonal ladders}. 
In the large-block-size limit, $x_n$ approaches to the value 
\begin{equation}
x \equiv \lim_{n \to \infty} x_n =\frac{3}{40+\sqrt{1627}}
\end{equation}
and hence we obtain the von Neumann entropy 
\begin{equation}
{\cal S}^{\rm 2leg}_\infty = -\sum^4_{i=1} p_i (\infty) \ln p_i(\infty) = 1.3781854.
\end{equation}
The von Neumann entropy per valence bond is given by ${\cal S}^{\rm 2leg}_\infty/2=0.6890927$, which is strictly less than $\ln 2$. 
\begin{table}[h]
\begin{center}
\vspace{0.1in}
\begin{tabular}{c|c|c|c|c|c} \hline
   & $N_x=1$ & $N_x=2$ & $N_x=3$ & $N_x=4$ & $N_x=5$ \\ 
\hline
$\shelf{N_y=4}{\rm Exact}{\rm MC}$ & $\vectorize{0.6891577}{0.6891575}$ & $\vectorize{0.6890932}{0.6890924}$ & 
$\vectorize{0.6890927}{0.6890929}$& $\vectorize{0.6890927}{0.6890925}$ & $\vectorize{0.6890927}{0.6890840}$\\
\hline
$\shelf{N_y=6}{\rm Exact}{\rm MC}$ & $\vectorize{0.6878024}{0.6878027}$ & $\vectorize{0.6876554}{0.6876558}$ & $\vectorize{0.6876523}{0.6876513}$ & $\vectorize{0.6876522}{0.6876537}$ & $\vectorize{0.6876522}{0.6875899}$ \\
\hline
$\shelf{N_y=8}{\rm Exact}{\rm MC}$ & $\vectorize{0.6871245}{0.6871243}$ & $\vectorize{0.6869344}{0.6869385}$ & $\vectorize{0.6869295}{0.6869363}$ & $\vectorize{0.6869293}{0.6868834}$ & $\vectorize{0.6869293}{0.6867750}$ \\
\hline
\end{tabular}
\caption{${\cal S}/|\Lambda_A|$ with $|\Lambda_A|=N_y/2$ for various $N_x$ obtained by algebraic (Exact) and numerical (MC) methods. The numerical values are rounded off to 8 decimal places.} 
\label{tab: hexagonal ladders} 
\end{center}
\end{table}

\subsubsection{3-leg and 4-leg hexagonal ladder}
\begin{figure}[htb]
\begin{center}
\vspace{.5cm}
\hspace{-.0cm}\includegraphics[width=0.6\columnwidth]{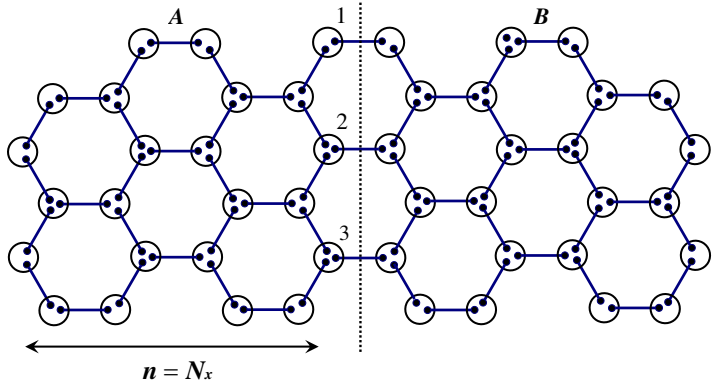}
\vspace{.0cm}
\caption{VBS state on a 3-leg hexagonal ladder. 
}
\label{fig: hexagonal_VBS_ladder2}
\end{center}
\end{figure}

Next we shall study the VBS states on the 3-leg and 4-leg hexagonal ladders.
As an illustration, in Fig. \ref{fig: hexagonal_VBS_ladder2}, we show an example of 3-leg hexagonal ladders. 
We first study the 3-leg ladders. In this case, the $Z$ matrix is defined similarly to Eq. (\ref{Z_n_3leg}):
\begin{equation}
Z_n(\vec \sigma_1, \vec \sigma_2, \vec \sigma_3)=
{\ssf a}_n+{\ssf b}_n \vec \sigma_1 \cdot \vec \sigma_2+{\ssf c}_n \vec \sigma_2 \cdot \vec \sigma_3+ {\ssf d}_n \vec \sigma_1 \cdot \vec \sigma_3,
\label{Z_n_3leg_h}
\end{equation}
with ${\ssf a}_0=1$ and ${\ssf b}_0={\ssf c}_0={\ssf d}_0=0$. The recursion relation becomes slightly complicated compared to the square cases due to the wiggly nature of the hexagonal ladders. It is composed of 2 steps as follows:
\begin{eqnarray}
({\ssf a}_{2m+1}, {\ssf b}_{2m+1}, {\ssf c}_{2m+1}, {\ssf d}_{2m+1})^{\rm T} &= T_{1} T_{2} ({\ssf a}_{2m-1}, {\ssf b}_{2m-1}, {\ssf c}_{2m-1}, {\ssf d}_{2m-1})^{\rm T}, \\
({\ssf a}_{2m+2}, {\ssf b}_{2m+2}, {\ssf c}_{2m+2}, {\ssf d}_{2m+2})^{\rm T} &= T_{2} T_{1} ({\ssf a}_{2m}, {\ssf b}_{2m}, {\ssf c}_{2m}, {\ssf d}_{2m})^{\rm T},
\end{eqnarray}
where two matrices $T_1$ and $T_2$ are given by
\begin{equation}
T_1=
\left(\begin{array}{cccc} 
1 & q^2 & q^2 & q^4 \\
q^3 & q^4 & q^5 & q^6 \\
q^3 & q^4 & q^4 & q^4 \\
q^5 & q^4 & q^6 & q^4
\end{array}\right),~~~~
T_2=
\left(\begin{array}{cccc} 
 1  & q^2 & q^2 & q^4 \\
q^3 & q^4 & q^4 & q^4 \\
q^3 & q^5 & q^4 & q^6 \\
q^5 & q^6 & q^4 & q^4
\end{array}\right),
\end{equation}
respectively, with $q=1/3$. Using the above recursion relation, one can obtain the coefficients $\{ {\ssf a}_n, {\ssf b}_n, {\ssf c}_n, {\ssf d}_n \}$. By diagonalizing the matrix in Eq. (\ref{Z_n_3leg_h}) with these coefficients, we can calculate the von Neumann entropy along the same lines as Sec. \ref{sec: 3-leg square ladder}. The obtained results are summarized in Table \ref{tab: hexagonal ladders}, which are compared with numerical results obtained in Sec. \ref{sec: MC hexagonal}. We now consider the large-block-size limit. Let $\vec v_{\rm PF}$ be the Perron-Frobenius vector of $T_1 T_2$. 
This vector is numerically given by
\begin{equation}
\vec v_{\rm PF}=(1,~0.03734899,~0.03734899,~0.0046503105)^{\rm T}
\end{equation}
Using the above values, we obtain the von Neumann entropy in the large-block-size limit as
\begin{equation}
{\cal S}^{\rm 3leg}_{\infty}=2.0629566,
\end{equation}
and hence the von Neumann entropy per valence bond is given by ${\cal S}^{\rm 3leg}_{\infty}/3=0.6876522$, which is strictly less than $\ln 2$. 

As is obvious, the argument so far can be straightforwardly generalized to $m$-leg ladders with $m \ge 4$~\cite{next_paper}. In Table \ref{tab: hexagonal ladders}, we show the results for the 4-leg hexagonal ladders. 
All the analytically obtained results indicate that the boundary correlation does not vanish even in the large-block-size limit and support the conjecture in Sec. \ref{sec: schmidt}, i.e., the eigenvalues of the reduced density matrix can be different. 
A numerical implementation of the transfer matrix technique 
for the $m$-leg ladders with $m > 4$ will be discussed elsewhere~\cite{next_paper}.

\section{Vertical ladders}
\begin{figure}[htb]
\begin{center}
\vspace{.5cm}
\hspace{-.0cm}\includegraphics[width=0.6\columnwidth]{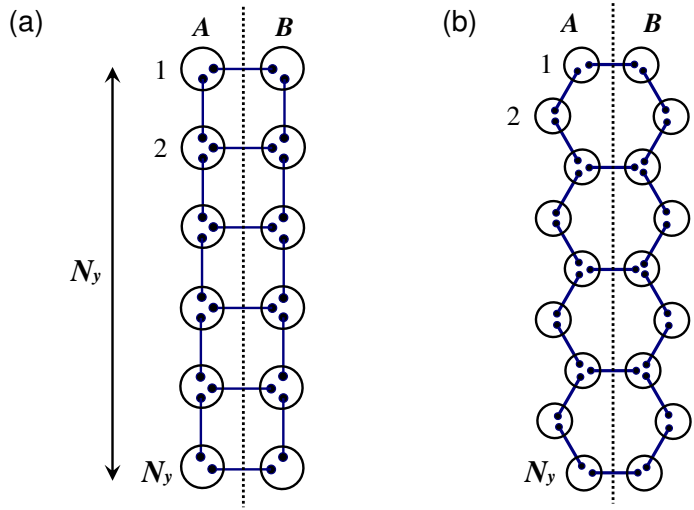}
\vspace{.0cm}
\caption{(a) Square lattice vertical ladder. (b) Hexagonal lattice vertical ladder. }
\label{fig: vertical_ladder}
\end{center}
\end{figure}

\begin{figure}[htb]
\begin{center}
\vspace{.5cm}
\hspace{-.0cm}\includegraphics[width=0.6\columnwidth]{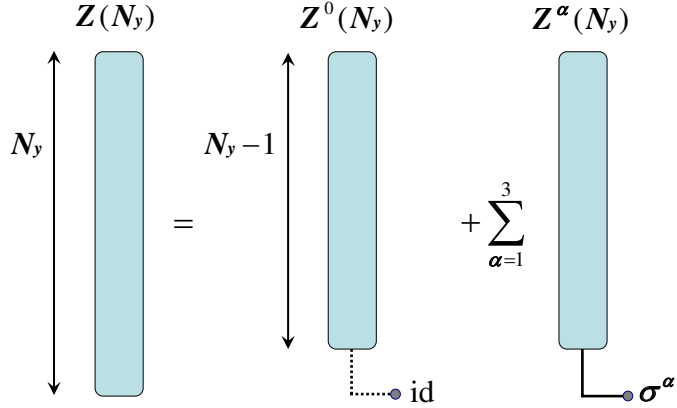}
\vspace{.0cm}
\caption{Decomposition of the matrix $N(N_y)$.}
\label{fig: decomp}
\end{center}
\end{figure}

\begin{figure}[htb]
\begin{center}
$\begin{array}{ccc}
\includegraphics[width=0.45\columnwidth]{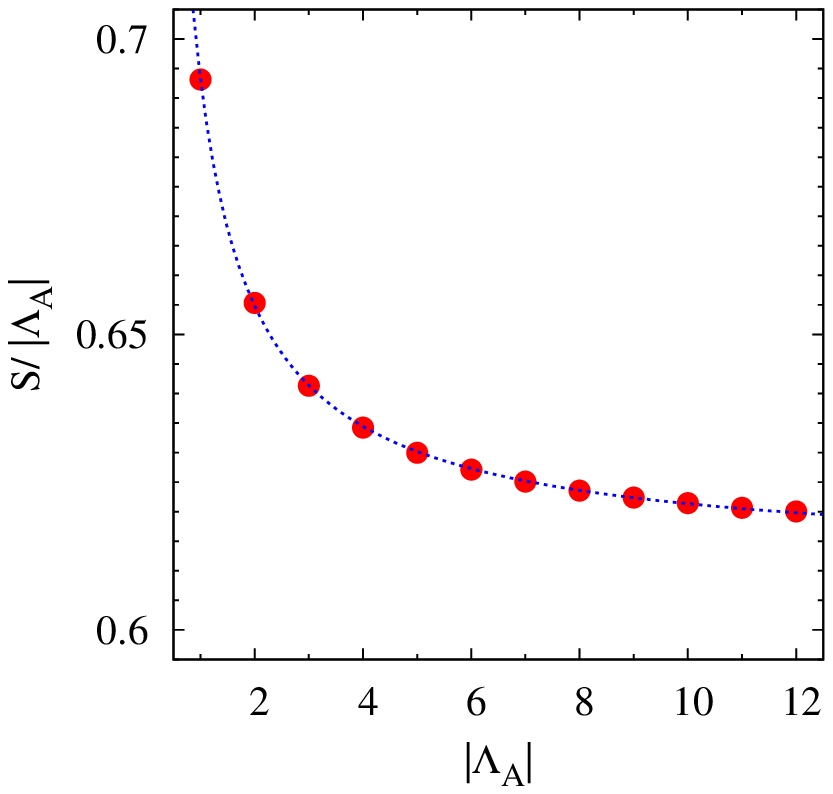}
& \hspace{5mm} &
\includegraphics[width=0.45\columnwidth]{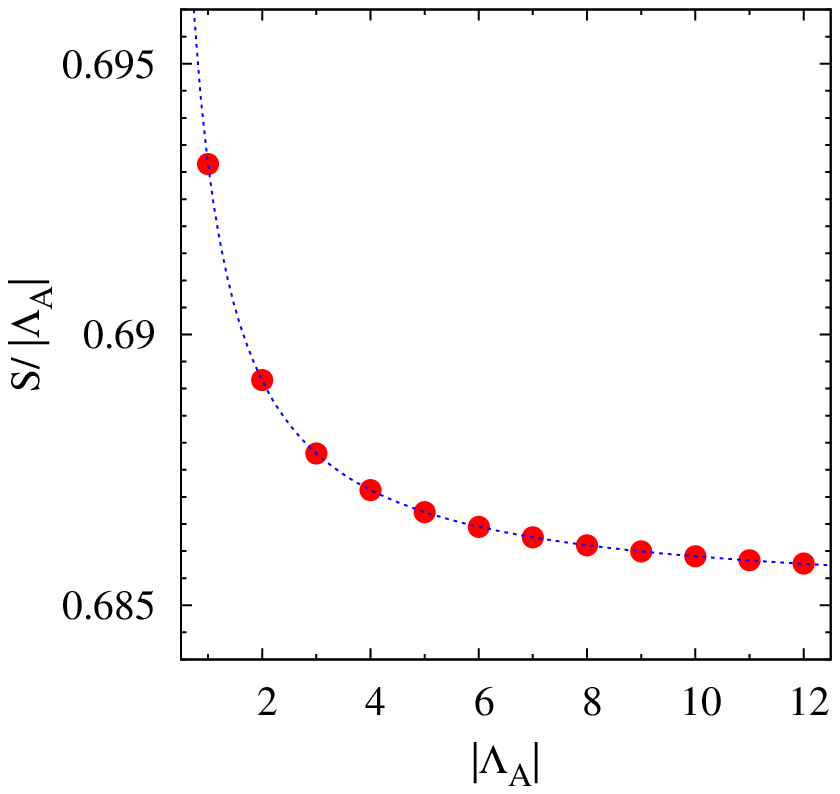} \\
({\rm a}) 
& \hspace{5mm} &
({\rm b})
\end{array}$
\caption{(a) The von Neumann entropy per valence bond as a function of $N_y$ for the square lattice VBS state with $N_x=1$. The blue curve shows the fitting curve (Eq. (\ref{eq:fitting})) with 
$C$ = 0.0819(3), 
$\Delta$ = 0.91(1), 
$\alpha$ = 0.6113(3).
(b) The von Neumann entropy per valence bond as a function of $N_y$ for the hexagonal lattice VBS state with $N_x$ = 1. The blue curve shows the fitting curve (Eq. (\ref{eq:fitting})) with
$C$ = 0.008081(5), 
$\Delta$ = 0.985(1), 
$\alpha$ = 0.685068(5).
}
\label{fig: algebraic_nx1}
\end{center}
\end{figure}

So far, we have studied square or hexagonal ladders in a horizontal geometry. In this section, we shall study the opposite limit, i.e., the ladders in a vertical geometry. Let us first consider the square ladder with $N_x=1$ and the height $N_y$ shown in Fig. \ref{fig: vertical_ladder}(a).
\\
Then the $Z$ matrix is given by
\begin{equation}
\fl~~~~~~~~~~
Z(N_y;\vec \sigma_1, ..., \vec \sigma_{N_y}) = \int \left( \prod^{N_y}_{i=1} \frac{d{\hat \Omega}_i}{4 \pi}  \right)
\prod^{N_y}_{j=1} ({1+ {\hat \Omega}_j \cdot \vec \sigma_j} )
\prod^{N_y-1}_{k=1} ( {1-{\hat \Omega}_k \cdot {\hat \Omega}_{k+1}}). 
\label{N mat 1}
\end{equation}
Similarly to the horizontal ladder cases, we can use the integral formula Eq. (\ref{3.3}) and rewrite Eq. (\ref{N mat 1}) as
\begin{equation}
\fl 
Z(N_y)= \sum_{\{ i_1, ..., i_{2n} \}} q^{n} (-q)^{i_2-i_1} (-q)^{i_4-i_3} \dots (-q)^{i_{2n}-i_{2n-1}}
(\vec \sigma_{i_1} \cdot \vec \sigma_{i_2}) (\vec \sigma_{i_3} \cdot \vec \sigma_{i_4}) \dots
(\vec \sigma_{i_{2n-1}} \cdot \vec \sigma_{i_{2n}}),
\end{equation}
where $1 \le i_1 < i_2 < \dots < i_{2n-1} < i_{2n} \le N_y$. 
Here we have abbreviated $Z(N_y)=Z(N_y;\vec \sigma_1, ..., \vec \sigma_{N_y})$.
The above formula can also be obtained using loops and strands shown in \ref{sec:loop model}. Note that $Z(1)$ is $2 \times 2$ identity matrix. We now look for a recursion relation between $Z(N_y+1)$ and $Z(N_y)$. 
We first decompose $Z(N_y)$ into four matrices as 
\begin{equation}
Z(N_y)= Z^0(N_y) + \sum^3_{\alpha=1} Z^\alpha(N_y),
\end{equation}
where $Z^0 (N_y)$ trivially acts on the $N_y$-th vector space $V_{N_y}$ while $Z^\alpha (N_y)$ acts on it as $\sigma^\alpha$. This decomposition is graphically shown in Fig. \ref{fig: decomp}. 
Using the graphical representation, we find the following set of recursion relations:
\begin{eqnarray}
\fl~~ 
Z^0 (N_y+1) &=& Z(N_y) \otimes {\rm id},\\
\fl~~ 
Z^\alpha (N_y+1) &=& -q^{2} Z(N_y-1) \otimes \sigma^\alpha \otimes \sigma^\alpha 
-q \left[ Z^\alpha(N_y). \left( \bigotimes^{N_y-1}_{j=1}{\rm id} \right) \otimes \sigma^\alpha \right] \otimes \sigma^\alpha, \\
\fl~~ 
Z(N_y+1) &=& Z^0 (N_y+1) +\sum^3_{\alpha=1} N^\alpha(N_y+1),
\end{eqnarray}
where `.' denotes matrix multiplication while `$\otimes$' denotes tensor product, and `${\rm id}$' is the $2 \times 2$ identity matrix. Solving the above recursions by {\it mathematica}, we obtain the overlap matrix $Z(N_y)$ exactly up to $N_y=12$. 
Then we numerically diagonalize $Z(N_y)$ and calculate the von Neumann entropy as a function of $N_y$. Figure \ref{fig: algebraic_nx1}(a) shows the result. The obtained values and also fitting parameters in Eq. (\ref{eq:fitting}) show good agreement with the Monte Carlo results. 

Let us next consider the hexagonal ladder with $N_x=1$ and the height $N_y$ shown in Fig. \ref{fig: vertical_ladder}(b). 
It is convenient to set $N_y=2m-1$ since $N_y$ is always odd in this case. 
In this case, from Eq. (\ref{eq: def of Z}), the $Z$ matrix is given by
\bege
\fl~~~~~~~~~~
Z(m; \vec \sigma_1, ..., \vec \sigma_m)=
\int \left( \prod^{N_y}_{i=1}\frac{d{\Oh}_i}{4\pi} \right)
\prod^m_{j=1}(1+\Oh_{2j-1}\cdot \vec \sigma_j)
\prod^{N_y-1}_{k=1}(1-\Oh_k \cdot \Oh_{k+1}).
\ende
Along the same lines as the square ladder case, we obtain the set of recursion relations for the $Z$ matrix:
\begin{eqnarray}
\fl~~
Z^0(m+1)= Z(m) \otimes {\rm id}, \\
\fl~~
Z^\alpha (m+1)=q^3 Z(m-1) \otimes \sigma^\alpha \otimes \sigma^\alpha
+q^2 \left[ Z^\alpha (m). \left( \bigotimes^{m-1}_{j=1}{\rm id} \right) \otimes \sigma^\alpha \right] \otimes \sigma^\alpha, \\
\fl~~
Z(m+1)=Z^0(m+1)+\sum^3_{\alpha=1}Z^\alpha (m+1).
\end{eqnarray}
Here we have abbreviated $Z(m)=Z(m; \vec \sigma_1, ..., \vec \sigma_m)$. 
Again, solving the above recursions, we find the $Z$ matrix up to $m=12$ ($N_y=23$). The numerically calculated von Neumann entropy by this method is shown in Fig. \ref{fig: algebraic_nx1}(b). The obtained values and the fitting parameters also show good agreement with the Monte Carlo results. 
The obtained results for square and hexagonal ladders with $N_x=1$ strongly support that the assumed fitting function Eq. (\ref{eq:fitting}) correctly describes the behavior of large-$N_y$ systems. 

\section{Conclusion}
\label{sec:}
In conclusion, we have studied entanglement properties of the VBS states on two-dimensional graphs with reflection symmetry. 
We have shown that the reflection symmetry permits us to develop an efficient method to study the reduced density matrix of the subsystem which is a mirror image of the other one. 
We found the relation between the reduced density matrix and the overlap matrix whose element is given by the inner product between degenerate ground states of the block Hamiltonian defined in the subsystem. 
This relation enables us to conjecture that the von Neumann entropy per valence bond on the boundary between the subsystems is strictly less than $\ln 2$ in 2d systems. 
This conjecture means that the entanglement spectrum is not flat in this system and is in contrast to the case of 1d VBS where the reduced density matrix is proportional to the identity matrix in the large-block limit. 
We have given a {\it holographic} interpretation of the reduced density matrix in terms of the spin chain along the boundary between the subsystems, which describes a hidden correlation among the degenerate ground states of the block Hamiltonian. 
To confirm the conjecture, we have numerically studied the eigenvalues of the reduced density matrix for finite square and hexagonal graphs by combining Monte Carlo integration with exact diagonalization. 
From the eigenvalues of the reduced density matrix, we have calculated the von Neumann entropy and have found that the von Neumann entropy per valence bond is well fitted by a constant term with a function decaying algebraically in the length of the boundary. The constant is strictly less than $\ln 2$ and supports our conjecture. We have also analytically and algebraically studied the quasi-1d cases where graphs are on ladders. The analytical results are totally consistent with the numerical results and strongly support our conjecture. Although our conjecture is very plausible, we still lack a rigorous proof of it in the infinite 2d system. Therefore, it would of course be interesting to study the problem by a completely different approach based on inequalities and perhaps reflection positivity. 

\ack{}
The authors are grateful to Zhengcheng Gu, Akimasa Miyake, Masaki Oshikawa, Yutaka Shikano, Xiaoliang Qi, and Ying Xu for their valuable comments and discussions. The creative atmosphere of Erwin Schr{\"o}dinger Institute was
also important for advancing the work. 
The present work was supported in part by NSF Grant DMS 0905744, by MEXT Grant-in-Aid for Scientific Research (B) (19340109) and for Scientific
Research on Priority Areas ``Novel States of Matter Induced by
Frustration'' (19052004), and by Next Generation Supercomputing Project,
Nanoscience Program, MEXT, Japan. 
H.K. is partly supported by the JSPS Postdoctral Fellow for Research Abroad.
S.T. is partly supported by Grant-in-Aid for Young Scientists Start-up (21840021) from the JSPS.
The computation is executed on computers at the Supercomputer Center, ISSP.

\appendix
\section{Graphical interpretation of $Z$ matrix}
\label{sec:loop model}
\begin{figure}[tb]
\begin{center}
\vspace{.5cm}
\hspace{-.0cm}\includegraphics[width=0.7\columnwidth]{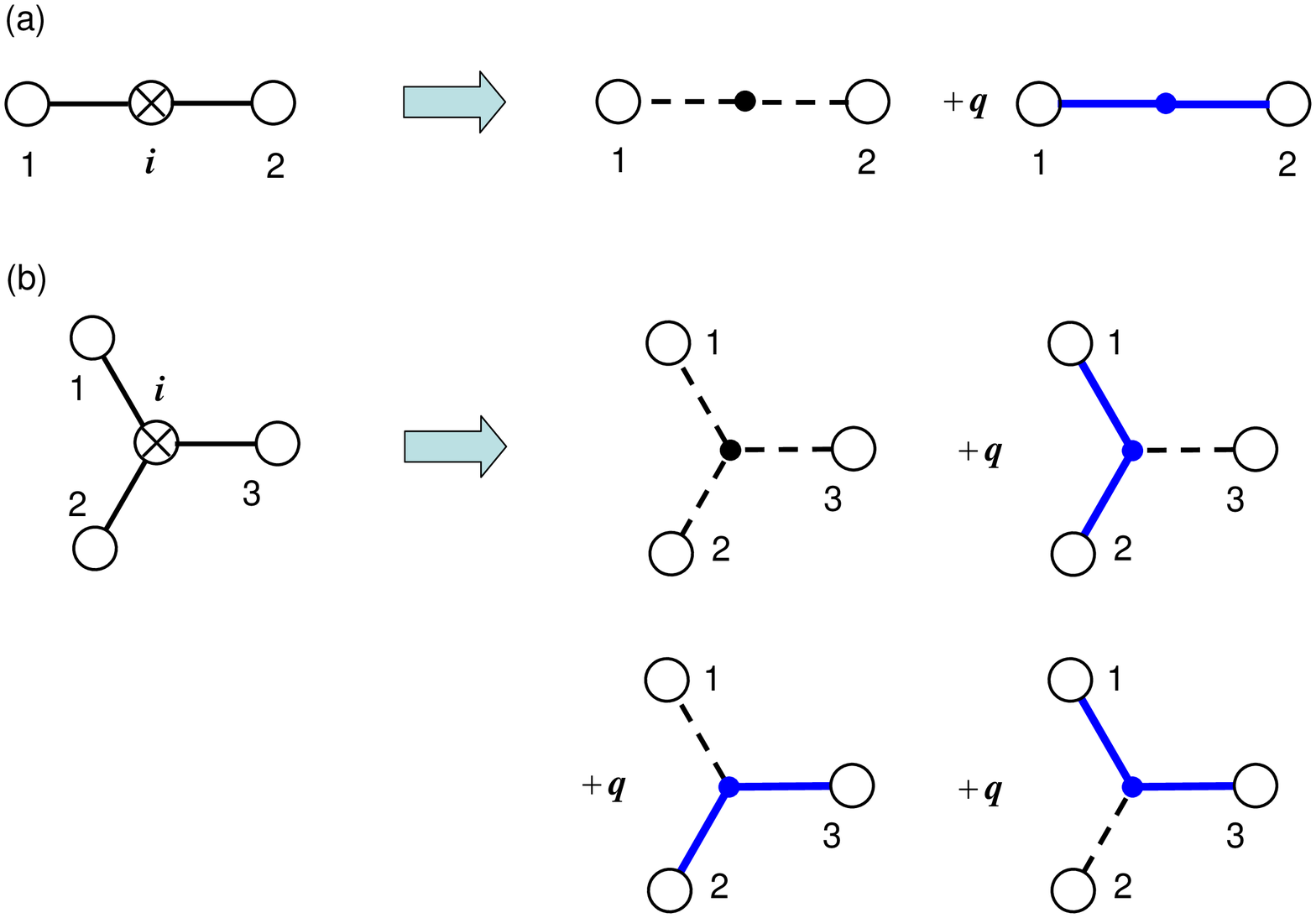}
\vspace{.0cm}
\caption{(a) Graphical representation for Eq. (\ref{loop1}).
The mark $\otimes$ denotes the integration over the spherical angle $\Oh_i$.
The broken and colored (thick) lines have weight $1$ and $q=1/3$, respectively.  (b)
Graphical representation for Eq. (\ref{loop2}). Notations are the same as (a).}
\label{fig: KLT}
\end{center}
\end{figure}

\begin{figure}[tb]
\begin{center}
\vspace{.5cm}
\hspace{-.0cm}\includegraphics[width=0.8\columnwidth]{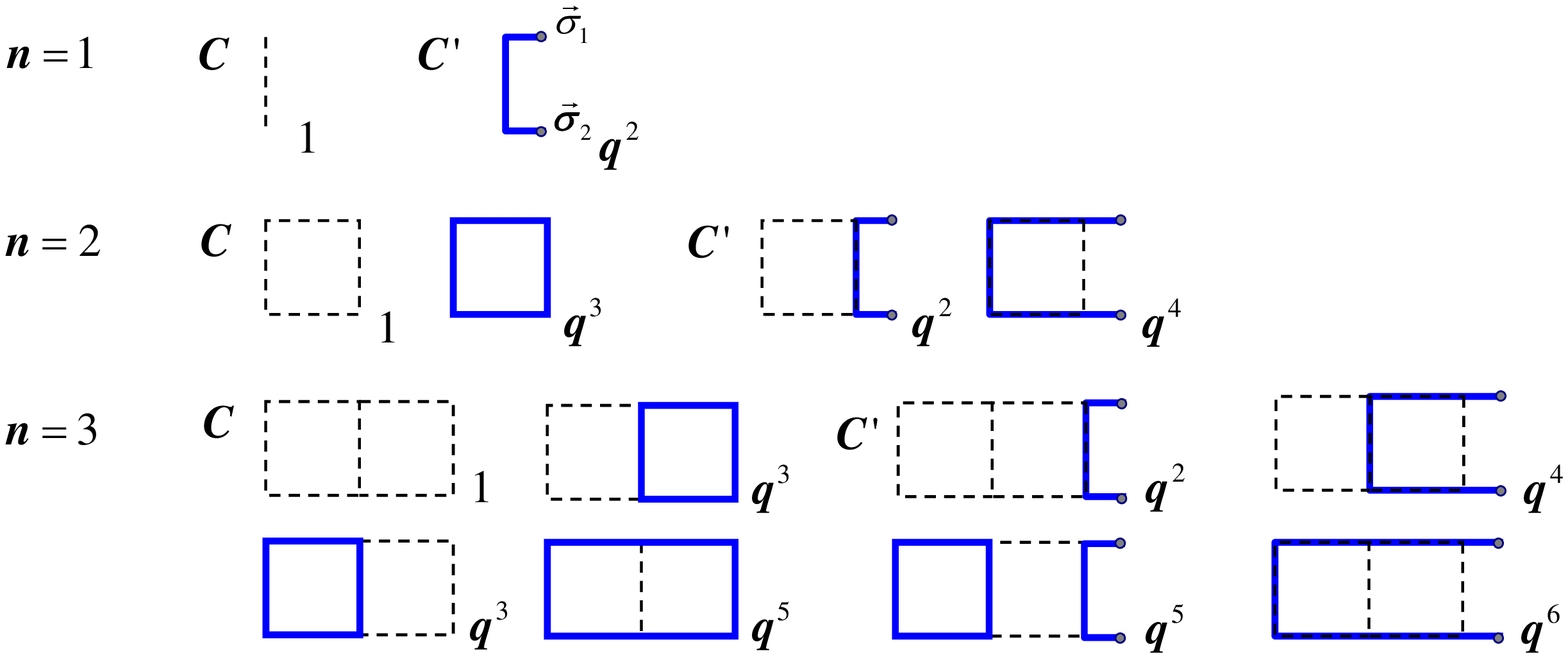}
\vspace{.0cm}
\caption{Closed loop ${\cal C}$ and open strand ${\cal C}'$ configurations for the length $n=1,2$, and $3$.
The weight of each configuration is assigned at the bottom right of each figure.}
\label{fig: config}
\end{center}
\end{figure}

In this appendix, we illustrate a graphical interpretation of the overlap matrix $Z$ (see Eq. (\ref{eq: integral=2leg_square})) in terms of loops and strands. 
As we have already mentioned in Sec. \ref{sec: square ladders}, the terms containing odd number of $\Oh_i$'s in
\begin{equation}
\prod_{(i,j)\in {\cal B}_A} (1-\Oh_i \cdot \Oh_j)=\sum_{\Gamma \subset {\cal B}_A} \prod_{(i,j)\in \Gamma} (-\Oh_i \cdot \Oh_j),
\end{equation}
vanish after integration over $\Oh_i$ 
since the measure of integration is invariant under the local change of variable, $\Oh_i \to -\Oh_i$. 
This can be graphically explained as follows. Suppose that one vertex $i \in A$ is connected to two vertices. They are labeled 1 and 2 (see Fig. \ref{fig: KLT}(a)).
Then we consider the following integral:
\begin{eqnarray}
  \int \frac{d\Oh_i}{4\pi} (1-\Oh_i \cdot \Oh_1)(1-\Oh_i \cdot \Oh_2)
= \int \frac{d\Oh_i}{4\pi} [1+(\Oh_i \cdot \Oh_1)(\Oh_i \cdot \Oh_2)].
\end{eqnarray} 
From the lemma 3.3 in Ref.~\cite{Kennedy_Lieb_Tasaki},
we can show
\begin{equation}
\int d\Oh_i (\Oh_1 \cdot \Oh_i) (\Oh_i \cdot \Oh_2)=\frac{4\pi}{3} \Oh_1 \cdot \Oh_2
\label{eq: integral_formula_1}
\end{equation}
and hence we obtain
\begin{equation}
\int \frac{d\Oh_i}{4\pi} (1-\Oh_i \cdot \Oh_1)(1-\Oh_i \cdot \Oh_2)
=1+q\, \Oh_1 \cdot \Oh_2,
\label{loop1}
\end{equation}
where $q=1/3$. This procedure is graphically shown in Fig. \ref{fig: KLT}(a).
Next, suppose that one vertex $i$ is surrounded by three vertices labeled by 1, 2, and 3. Then we encounter with the following integral:
\begin{equation}
\int \frac{d\Oh_i}{4\pi} (1-\Oh_i \cdot \Oh_1)(1-\Oh_i \cdot \Oh_2)(1-\Oh_i \cdot \Oh_3).
\end{equation}
Using Eq. (\ref{eq: integral_formula_1}), again, we obtain
\begin{eqnarray}
&&  \int \frac{d\Oh_i}{4\pi} (1-\Oh_i \cdot \Oh_1)(1-\Oh_i \cdot \Oh_2)(1-\Oh_i \cdot \Oh_3) \nonumber \\
&=& \int \frac{d\Oh_i}{4\pi}[1+(\Oh_i \cdot \Oh_1)(\Oh_i \cdot \Oh_2)+(\Oh_i \cdot \Oh_2)(\Oh_i \cdot \Oh_3)+(\Oh_i \cdot \Oh_3)(\Oh_i \cdot \Oh_1)]
\nonumber \\
&=& 1+ q\, \Oh_1 \cdot \Oh_2 + q\, \Oh_2 \cdot \Oh_3 + q\, \Oh_3 \cdot \Oh_1.
\label{loop2}
\end{eqnarray}
This procedure can also be graphically interpreted as shown in Fig. \ref{fig: KLT}(b). 
Therefore, as we have seen, the integrand containing odd number of $\Oh_i$'s will vanish. In terms of the graphical representation, the matrix $Z_n$ for the 2-leg square ladder can be written as
\begin{eqnarray}
Z_n(\vec \sigma_1, \vec \sigma_2) = \sum_{\{{\cal C} \}} q^{N_{\rm B}({\cal C})} q^{-N_L({\cal C})}
-\sum_{\{{\cal C}' \}} q^{1+N_{\rm B}( \{\cal C' \}) }q^{-N_{\rm L}(\{ {\cal C}' \})} ({\vec \sigma}_1 \cdot {\vec \sigma_2}),
\end{eqnarray}
where $N_{\rm B}({\cal C})$ and $N_{\rm L}({\cal C})$ denote the numbers of colored edges (bonds) and loops in the configuration ${\cal C}$, respectively. 
The minus sign stems from the product like $(1+\Oh_1 \cdot {\vec \sigma}_1)(1-\Oh_1 \cdot \Oh_2)(1+\Oh_2 \cdot {\vec \sigma}_2)$.
In the first sum over $\{ {\cal C}\}$, only the closed loops are allowed.
On the other hand, in the second sum over $\{{\cal C}' \}$, open ends from $1$ and $2$ are allowed (see Fig. \ref{fig: config}).
Therefore, we can obtain the coefficients ${\ssf a}_n$ and ${\ssf b}_n$ in Eq. (\ref{eq: Zmatrix_2leg_square}) by a completely graphical or combinatorial way.  
From Fig. \ref{fig: config}, one can read ${\ssf a}_1=1$, ${\ssf a}_2=1+q^{3}$, ${\ssf a}_3=1+2q^{3}+q^{5}$ and ${\ssf b}_1=q^{2}$, ${\ssf b}_2=q^{2}+q^{4}$, ${\ssf b}_3=q^{2}+q^{4}+q^{5}+q^{6}$. 
Although we only show the case of 2-leg square ladders for simplicity, one can easily generalize this approach to any AKLT ladders.

\section*{References}
\providecommand{\newblock}{}

\end{document}